# Programmable Multi-responsive Nanocellulose-based Hydrogels with Embodied Logic

*Beatriz Arsuffi* [1, 2, 3], *Gilberto Siqueira*[2] *, *Gustav Nyström*[2], *Silvia Titotto*[3] *,

*Tommaso Magrini*[1, 4] *, *Chiara Daraio*[1]*

1 - Department of Mechanical and Civil Engineering, California Institute of Technology, Pasadena, CA 91125, USA.

2 - Empa, Swiss Federal Laboratories for Materials Science and Technology, Cellulose and Wood Materials Laboratory, 8600, Dübendorf, Switzerland.

3 - Center of Engineering, Modelling and Applied Social Sciences, Federal University of ABC, Santo André, SP, Brazil.

4 - Department of Mechanical Engineering, Eindhoven University of Technology, 5600MB Eindhoven, The Netherlands.

* Email: gilberto.siqueira@empa.ch, silvia.titotto@ufabc.edu.br, t.magrini@tue.nl, daraio@caltech.edu

Programmable materials are desirable for a variety of functional applications that range from biomedical devices, actuators and soft robots to adaptive surfaces and deployable structures. However, current smart materials are often designed to respond to single stimuli (like temperature, humidity, or light). Here, a novel multi-stimuli-responsive composite is fabricated using direct ink writing (DIW) to enable programmability in both space and time and computation of logic operations. The composite hydrogels consist of double-network matrices of poly(*N*-isopropylacrylamide) (PNIPAM) or poly(acrylic acid) (PAA) and sodium alginate (SA) and are reinforced by a high content of cellulose nanocrystals (CNC) (14 wt%) and nanofibers (CNF) (1 wt%). These composites exhibit a simultaneously tunable response to external stimuli, such as temperature, pH, and ion concentration, enabling precise control over their swelling and shrinking behavior, shape, and mechanical properties over time. Bilayer hydrogel actuators are designed to display bidirectional bending in response to various stimuli scenarios. Finally, to leverage the multi-responsiveness and programmability of this new composite, Boolean algebra concepts are used to design and execute NOT, YES, OR, and AND logic gates, paving the way for self-actuating materials with embodied logic.

## 1. Introduction

Programmable materials are a novel class of materials that can self-actuate in response to external triggers, such as temperature, variations in pH, light intensity, as well as chemical, electric and magnetic fields. Such actuation must occur in a predictable manner, through controlled and gradual changes in physical and/or chemical properties. [1,2] In this context, responsive materials that can achieve a programmable shape transformation play a crucial role in the advancement of various fields including biomedical devices [3], microfluidics [4], and soft robotics. [5] However, while natural environments offer multiple stimuli at the same time, smart materials are often limited to respond to only one or two stimuli. [6,7] Furthermore, smart materials frequently lack precise temporal control over actuation, necessary for their full implementation for end-user applications. [8] Unlike smart materials, the design of programmable materials extends further than the conventional characterization of material properties. For example, if-then relationships, feedback mechanisms, information processing tools, and logic operations can be used to introduce varying degrees of intelligence into the material system, enabling the programmability and control of its entire behavior through adaptable strategies. [9–11] Materials with embodied logic generate predictable outputs in response to mechanical or chemical inputs, enabling materials-driven computation tailored to the specific physics and timescales of the target application. [12–15]

Hydrogels are among the most promising programmable materials that are used for conducting logic operations. [16–18] Hydrogels are hydrophilic polymeric materials with crosslinks between the chains, forming a three-dimensional network structure. [19,20] In contact with water or aqueous solutions, hydrogels can absorb large amounts of water, swelling hundreds of times their dry polymer network mass. [21] The most common responsive hydrogels are triggered by heat and are characterized by the presence of hydrophobic groups, such as methyl, ethyl, and propyl. [22] Poly($N$-isopropylacrylamide) (PNIPAM) is one of the most widely employed thermoresponsive hydrogels, particularly for biomedical applications, since its volume phase transition temperature (VPTT) around 32°C is close the body temperature and can be by copolymerization with other co-monomers. [23–25] pH-responsive hydrogels are also extensively utilized as programmable soft materials. In these hydrogels, hydrophilic networks undergo volumetric changes in reaction to variations in the surrounding pH levels. [26] The fundamental components of such hydrogels are polymers with weakly acidic or basic properties, such as poly(acrylic acid) (PAA). [27] In addition, ions play diverse roles in inducing changes in hydrogel swelling, depending on the hydrogel's chemical structure. [28] For instance, ionic interactions have been widely employed to establish physical crosslinking in hydrogels with charged components, such as sodium alginate (SA). [29,30]

Nonetheless, hydrogels commonly exhibit poor mechanical and structural properties, limiting their actuation power and applications. [31,32] Strategies to address this limitation include synthesizing double-network hydrogels[33], or incorporating reinforcing particles such as nanocelluloses (NC). [34,35] When successfully aligned during hydrogel fabrication, NC reinforcing particles can impart an anisotropic swelling behavior, as well as superior mechanical properties. [36,37] For instance, PNIPAM hydrogels reinforced with 20 wt% CNC presented an increase in stiffness by a factor of 236 compared to the pure matrix. [38] One method to achieve high NC alignment is through direct ink writing (DIW) 3D printing technique. [35,39] In DIW, NC-laden inks are extruded through a nozzle, achieving precise control of the NC particles alignment, critical to achieve the intended in plane actuation, that is necessary for volumetric shape-morphing. [39] The anisotropic shape transformations induced by NC alignment have been used to fabricate programmable actuators within the realm of 4D printing. [40,41] 4D printing consists of prototyping responsive and time-dependent structures through 3D printing technologies. [42] The fourth dimension refers to the capability of the structure to alter its shape, functionality, and/or properties over time in response to specific environmental stimuli during its post-printing lifetime. [43] Programmable hydrogels are among the most used materials for 4D printing [44], especially within architectures designed to enhance actuation, such as bilayer systems. The swelling mechanism driving bilayer morphism can be achieved through either multi- or mono-material printing. In the multi-material approach, varying water absorption capacities of each hydrogel cause the structure to bend towards the side with the hydrogel exhibiting lower swelling degree. [45,46] Whereas, in the mono-material approach, shape alteration arises from differences in the filament orientation within each layer. [47]

Here, we report a facile approach to manufacture multi-stimuli-responsive hydrogel structures capable of sensing three distinct environmental stimuli and computing logic operations through programmable mechanical actuation. In this approach, we employ five key strategies: (i) utilizing stimuli-responsive hydrogels: PNIPAM for temperature and ionic responses, and PAA for pH response; (ii) incorporating anisotropic reinforcing particles: cellulose nanocrystals (CNCs) and nanofibers (CNFs); (iii) designing the geometry of filament orientation; (iv) fabricating a bilayer system through multi-material 4D printing; and (v) tuning the stimuli concentration and combining different stimuli. We develop inks with a high nanocellulose content (15 wt%), investigate their printability, and adjust their processing parameters. Then, we characterize the swelling and shrinking behavior of the hydrogels, as well as their shape-morphing mechanism and mechanical properties in response to multiple stimuli excitations. Finally, we demonstrate the material's programmability both in space and time, exploring Boolean algebra concepts by designing logic gates that leverage the actuation of the bilayer structures.

## 2. Results and Discussion

### 2.1. Fabrication of Nanocellulose-based Multi-responsive Hydrogels

To fabricate multi-responsive programmable hydrogels, the first step involved preparation of the nanocelluloses. We extracted cellulose nanofibers (CNFs) from wood pulp through TEMPO-oxidation process **(Figure 1a)**. [48] Using mechanical mixing, we then combined CNFs, commercially available cellulose nanocrystals (CNCs), sodium alginate (SA), and the smart polymers poly(*N*-isopropylacrylamide) (PNIPAM) and poly(acrylic acid) (PAA) into a 3D printing ink (Figure 1b). PNIPAM is a widely known thermoresponsive hydrogel, characterized by a lower critical solution temperature (LCST) of approximately 32 °C. This implies that PNIPAM macromolecules undergo a volume phase transition (VPT) from a well-hydrated state (below 32 °C) to a collapsed state (above 32 °C) **(Figure 2a)**. [49] Thus, PNIPAM's volume is drastically altered by small changes in temperature due to the reversible transition of its polymer chains from a hydrophilic to a hydrophobic state at the LCST. [50] Additionally, salt solutions exceeding critical concentrations can also induce an analogous phase transition in PNIPAM. This is because the presence of anions in the surrounding media acts on reducing the LCST. Consequently, its swelling degree demonstrates a reversible decrease with an increase in ionic concentration. [51] In contrast to what is observed in PNIPAM-based hydrogels, the swelling behavior of PAA-based hydrogels is primarily influenced by the surrounding pH (Figure 2a). When the pH exceeds the pKa of 4.3, PAA-based hydrogels undergo swelling and they expand in volume. Conversely, when the pH falls below the pKa, the PAA-based hydrogel undergoes shrinking and decreases in volume. [52] As the PNIPAM- or PAA-hydrogel networks provide stimuli responsiveness, the secondary SA-based network adds rigidity and stiffness to our NC-reinforced double-network hydrogel system, drastically enhancing its structural properties (Figure S4, Supporting Information). [33,44]

The production of homogeneous inks for DIW with a high NC solid loading is non-trivial, yet it is key for high quality manufacturing, thus requiring the mastering of nanoparticles dispersion in aqueous or non-polar solvents. [35,53] Previous works that targeted similar hydrogel-based systems, have demonstrated that rheological properties such as shear-thinning behavior, fast elastic recovery, well-defined yield stress (in the order of 100 Pa), and elastic modulus higher than a few kPa are key parameters that ensure high shape fidelity during the DIW process. [35,38,39] In this work, we observed that these properties were achieved by incorporating CNCs (14 wt%) and CNFs (1 wt%) into the inks. Shear thinning was promoted by the alignment of the nanocellulose particles during extrusion (Figure 1c). [39] Yet, the flow-induced orientation of CNCs and CNFs occurs only if the applied stress during printing surpasses the ink's yield stress (i.e., the differential-flow regime). [35] Both CNC and CNF particles were essential for modifying the rheology of the inks and acting as anisotropic reinforcements in the final printed hydrogels (Section S1 and Figure S1, Supporting Information). This enabled the printing of complex-shaped structures with precise geometry control (Figure 1c), showcasing the fabrication versatility of the developed inks. High shape fidelity was achieved through a dual-step crosslinking process following 3D printing. Photopolymerization was employed to crosslink either the PNIPAM- or the PAA-based networks (Figure 1d), while ionic crosslinking with $Ca^{2+}$ was applied to the SA network (Figure 1c and Section S2, Supporting Information).

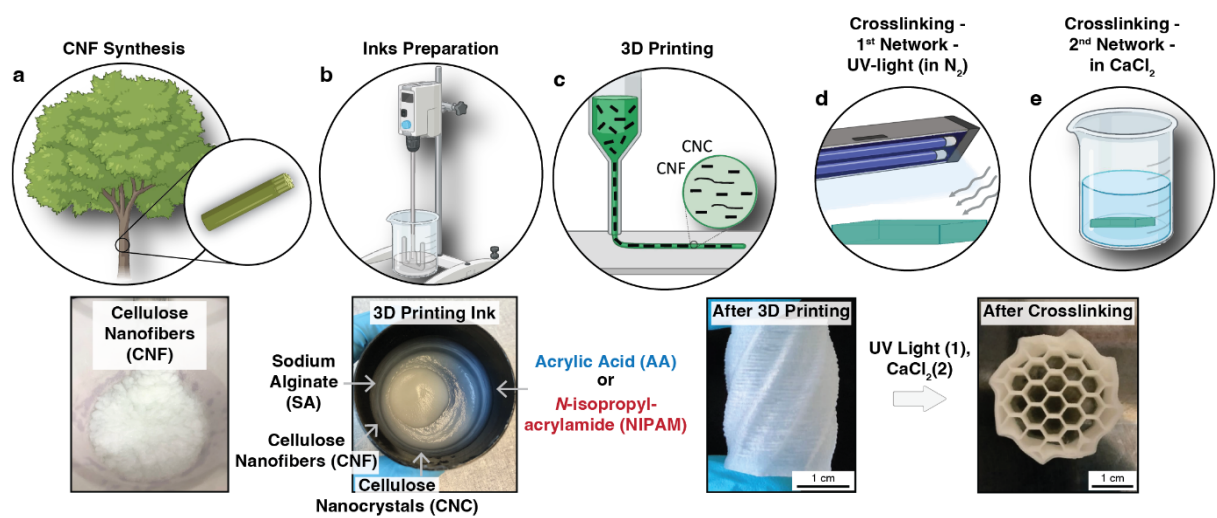

**Figure 1.** Fabrication process overview of multi-stimuli-responsive nanocellulose-based hydrogels: (A) Extraction of CNF from wood pulp. (B) Preparation of inks by mixing. (C) 3D printing via DIW. (D) Photopolymerization of the PNIPAM- or the PAA-based networks. (E) Ionic crosslinking of the SA network in $CaCl_2$.

## 2.2. Physical Properties of the Composites: Swelling and Shrinking Behavior, Shape-change, and Stiffness

The swelling capacity of the hydrogel composites was quantified through water absorption tests using fully dried 3D printed samples. Both PNIPAM/SA/NC and PAA/SA/NC hydrogels exhibit similar swelling kinetics, with a rapid growth in water uptake in the first 8 hours, followed by a prolonged stabilization, indicating a quick water absorption response due to the higher osmotic pressure between the dry hydrogels and water (Figure 2b). However, because PAA is a superabsorbent material, due to the deprotonation of the carboxylic acid groups of its polymer chain [52], PAA/SA/NC hydrogel shows much higher values of water absorption than PNIPAM/SA/NC hydrogel. While the PNIPAM/SA/NC hydrogel exhibits over 735% water absorption compared to its dry weight after 5 days of swelling, the PAA/SA/NC hydrogel presents more than three times higher water absorption (2499%) for the same period (Figure 2b). The equilibrium moisture content is achieved after 4 days, when the osmotic pressure balances the retractile forces exerted by the stretching polymer chains. [54] For both PNIPAM/SA/NC and PAA/SA/NC hydrogels, the swelling capacity can be adjusted by controlling the crosslinking density of the SA network. Increasing the $CaCl_2$ concentration, during crosslinking, reduces the hydrogels' water uptake. For example, the swelling capacity of PNIPAM/SA/NC hydrogel is reduced by a factor of 2.7 when crosslinked with 5 wt% $CaCl_2$, and PAA/SA/NC hydrogel not crosslinked with $CaCl_2$ has a water uptake 12.8 times higher than when crosslinked with 5 wt% $CaCl_2$ (Figure S3, Supporting Information). Higher $CaCl_2$ concentrations create a more extensively crosslinked SA network, restricting hydrogel expansion during swelling and increasing stiffness. Indeed, samples crosslinked with 5 wt% $CaCl_2$ exhibit a Young's modulus over 25 and 149 times higher than samples without ionic crosslinking for PNIPAM/SA/NC and PAA/SA/NC hydrogels, respectively (Figure S4, Supporting Information). This effect is intensified in PAA/SA/NC hydrogels due to their higher SA content compared to PNIPAM/SA/NC hydrogels.

As PNIPAM and PAA are stimuli-responsive hydrogels, the shrinking behavior of the composites upon exposure to environmental stimuli was studied using fully swollen 3D printed samples. When submerged in acidic solutions with a pH below PAA's pKa (4.3), the PAA/SA/NC hydrogel rapidly shrinks, reaching nearly maximal shrinkage in less than 30 minutes (Figure 2c). The shrinking rate of the PAA/SA/NC hydrogel increases as the pH level decreases (Section S3 and Figure S5, Supporting Information). For instance, after 1 hour in a solution with a pH of 2, the PAA/SA/NC hydrogel shows a water loss of 75% compared to its swollen state (Figure 2c). PNIPAM, in contrast, shrinks in response to temperatures above its

LCST (32 °C).[49] The shrinking degree of the PNIPAM/SA/NC hydrogel directly correlates with the heat intensity. At all tested temperatures (40 °C, 60 °C, and 80 °C), the shrinking kinetics of the PNIPAM/SA/NC hydrogel are faster than its swelling kinetics (Section S3 and Figure S5, Supporting Information). Within 1 hour at 80 °C, the PNIPAM/SA/NC hydrogel loses 54% of its water compared to its swollen state (Figure 2f). Moreover, the presence of anions in the surrounding media reduces the LCST of PNIPAM.[55] Therefore, the shrinking rate of PNIPAM/SA/NC hydrogel increases with higher salt concentrations in the solution (Section S3 and Figure S5, Supporting Information). For example, after 1 hour in a 6 M sodium chloride (NaCl) solution, the hydrogel shows a 32% reduction in water content compared to its swollen state (Figure 2f). Furthermore, when temperature and NaCl stimuli are combined, the shrinking behavior of PNIPAM/SA/NC hydrogel is accelerated. After just 5 minutes in a 6 M NaCl solution at 80 °C, the hydrogel presents a water loss value exceeding 40%, which increases to over 58% after 1 hour (Figure 2f). When the hydrogels shrink, they contract in size (Figure S6, Supporting Information). This shape oscillation can be programmed by adjusting the applied stimulus. A higher stimulus results in greater shrinking and a more pronounced shape change (Section S4 and Figure S8, Supporting Information). For instance, PAA/SA/NC hydrogel undergoes significant contraction in response to pH 2, reducing its top area by over 56% within 2 hours (Figure 2d). Similarly, the PNIPAM/SA/NC hydrogel's top area decreases by more than 48% at 80 °C and 43% in a 6 M NaCl solution after 2 hours. This contraction intensifies to 56% when the hydrogel is simultaneously exposed to both stimuli (Figure 2g).

The mechanical properties of the composites were assessed through compression tests on fully swollen 3D printed samples. As expected, the reinforcement with nanocellulose particles (both CNC and CNF) positively impacted the stiffness of the composites. The PNIPAM/SA/NC hydrogel has a Young's modulus of 252.94 kPa, which is 48 times higher than that of the PNIPAM/SA matrix without nanocellulose (Section S5 and Figure S9, Supporting Information). Moreover, when the smart hydrogels are exposed to external stimuli, they expel water and experience an increase in polymer chain entanglement, leading to an increase in stiffness.[22] In resonance with the shrinking behavior, the stiffness of the composites can be controlled by modulating the pH, temperature and salt concentration (Section S6 and Figure S10, Supporting Information). For instance, the PNIPAM/SA/NC hydrogel subjected to 80 °C for 24 hours shows a modulus 42% higher (323.54 kPa) than that of the swollen hydrogel at 21 °C (Figure 2h). This increase in stiffness is even more pronounced for samples immersed in a NaCl solution. The hydrogel previously immersed in 6 M NaCl exhibits a Young's modulus of 1782.69 kPa, nearly eight times higher than that of the hydrogel not exposed to NaCl (Figure 2h). However, the substantial increase in stiffness at high NaCl concentrations might also be influenced by salt aggregation within the printed structures, in addition to water loss. When both temperature (80 °C) and NaCl (6 M) stimuli are combined, the PNIPAM/SA/NC hydrogel demonstrates a Young's modulus nearly 15 times higher (3373.84 kPa) than in its swollen state (Figure 2h). A similar behavior is observed for the PAA/SA/NC hydrogel, which becomes stiffer with a decrease in pH, presenting Young's modulus values of 123.08 kPa at pH 7 and 178.55 kPa at pH 2 (Figure 2e). The PAA/SA/NC hydrogel exhibits more significant shrinking in response to pH variations compared to the shrinkage observed in the PNIPAM/SA/NC hydrogel when exposed to temperature and/or NaCl (Figure 2c,f). Despite this, the PAA/SA/NC hydrogel demonstrates a lower increase in Young's modulus during stimulus exposure compared to PNIPAM/SA/NC (Figure 2e,h). This difference can be attributed to the effects of $CaCl_2$ crosslinking. The PAA/SA/NC hydrogel contains twice as much SA as the PNIPAM/SA/NC hydrogel, and no $CaCl_2$ crosslinking was applied during the shrinking tests. Conversely, for the compression tests, the samples were pre-crosslinked with 5 wt% $CaCl_2$ for 24 hours.

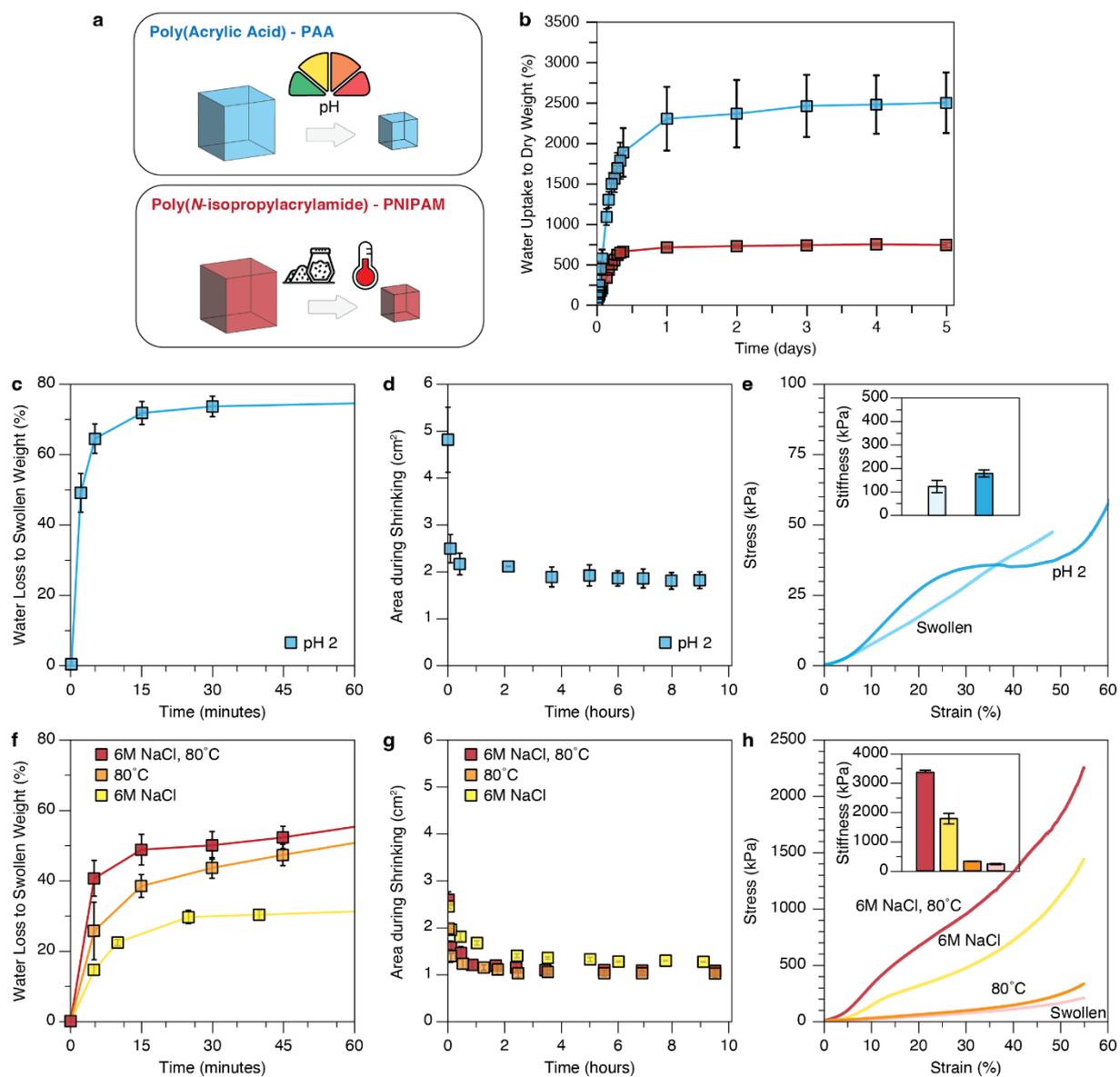

**Figure 2.** Physical properties of responsive hydrogels. (A) Responsivity of PAA- and PNIPAM-based hydrogels to external stimuli. (B) Swelling behavior of PNIPAM/SA/NC and PAA/SA/NC hydrogels. (C) Shrinking behavior of PAA/SA/NC hydrogel in response to pH. (D) Shape-change during shrinking of PAA/SA/NC hydrogel. (E) Comparison of average stress vs. strain curves and stiffness for PAA/SA/NC hydrogel in its swollen and shrunk states. (F) Shrinking behavior of PNIPAM/SA/NC hydrogel in response to temperature and/or salt concentration. (G) Shape-change during shrinking of PNIPAM/SA/NC hydrogel. (H) Comparison of average stress vs. strain curves and stiffness of PNIPAM/SA/NC hydrogel in its swollen and shrunk states. Error bars show standard deviation (n = 3).

## 2.3. Shape-morphing of Programmable Actuators

### 2.3.1. Programmability in Space: Anisotropic Shape-morphing

The intense shear and extensional forces applied during DIW induce the alignment of cellulose nanoparticles along the direction of the printed filaments. [39] This alignment facilitates controlled shape-morphing, as nanocellulose particles do not expand or contract along their axial direction. [56] Conversely, expansion or contraction primarily occurs perpendicular to the nanocellulose orientation. [38] Additionally, the alignment of cellulose nanoparticles induces different degrees of swelling/shrinking and internal tensions within the hydrogels structures, which can only be reduced through deformations [44], resulting in highly anisotropic actuation. The anisotropic shape-morphing capability of the PNIPAM/SA/NC hydrogel was investigated using bilayers printed with varying filament orientations (0°, 45°, and 90°), with both layers sharing the same orientation. In response to temperature (60 °C), the structures undergo significant and rapid actuation within seconds. As the hydrogel shrinks, the initially flat bilayers morph into well-defined 3D structures, following the predetermined filament orientation during 3D printing. Printing at 0° yields a curved structure, while printing at 90° and 45° produces rolled and twisted architectures, respectively (Figure S11, Supporting Information). In contrast, a cast sample, utilized for comparison, transforms randomly into an uncontrolled shape, underscoring that the shape transformations are guided by the alignment of cellulose nanoparticles (both CNC and CNF) within the printed structures. The actuation behavior of the developed materials is similar to the shape-shifting observed in plant systems, attributed to the reinforcement of cellulose microfibrils in cell walls, as seen in pinecone scales [57], wheat awns [58], and orchid tree seedpods. [59] Therefore, precise control over the morphology of the hydrogel structures at the microscale allows for programmable macroscopic shape-morphing akin to the morphological transformations observed in biological systems. This shape transformation can thus be spatially programmed and is reversible (Figure S11, Supporting Information). The PNIPAM/SA/NC bilayers, printed at 0°, can undergo 7 cycles of actuation, transitioning from DI water at 21 °C to 60 °C, with complete recovery of their original shape.

### 2.3.2. Programmability in Time: Shape-morphing of Multi-responsive Bilayers

The ability to control physical properties, such as shrinking capacity, shape, and stiffness, of the stimuli-responsive composites by adjusting their crosslinking density and stimuli intensity enables temporal programmability. This understanding of programmability served as the foundation for designing bilayer actuators with the PNIPAM/SA/NC and

PAA/SA/NC hydrogels endowed with controllable shape-morphing. As discussed in the previous section, the geometry of filament orientation significantly influences the shape-morphing of hydrogel sheets. To isolate the impact of this variable on shape alterations of the bilayers, both layers were printed at 0°. Moreover, strong interfacial bonding between the two layers is essential to prevent delamination caused by significant differential dimensional changes during swelling and shrinking. [60] This is especially important in our system since both PNIPAM and PAA contract in volume upon exposure to different stimuli. Furthermore, compatibility between the two hydrogels is crucial, considering factors such as material density, and geometric constraints. [44] Here, the adhesion between the layers was established through the polymeric network of SA, which is present in both layers, and the dual-step crosslinking process involving both materials simultaneously. Upon exposure to varying intensities of temperature, pH, and NaCl, the bilayers demonstrated shape-morphing behavior with bidirectional bending, corresponding to the actuation of either the PNIPAM/SA/NC or PAA/SA/NC layer.

In multi-material bilayer systems responsive to external stimuli, shape-morphing is driven by the differential swelling/shrinking capacities of each hydrogel, causing the structure to bend toward the side with the higher shrinking capacity. [45,46] For instance, when exposed to temperatures above 32 °C or high salt concentrations, the top layer of the designed bilayer system, composed of PNIPAM/SA/NC hydrogel, contracts, causing the structure to bend with PNIPAM/SA/NC as the inner layer. Conversely, when immersed in an acidic solution (pH < 4.3), the bottom layer consisting of PAA/SA/NC hydrogel contracts, resulting in bending in the opposite direction, with PAA/SA/NC as the inner layer **(Figure 3a,b)**. Thus, depending on the applied stimulus, one layer of the bilayer structure becomes activated and undergoes shrinking, while the other layer remains passive, facilitating bidirectional bending actuation. Moreover, this bending actuation can be programmed by tuning the condition of the surrounding environment. To achieve this, the bilayers were subjected to varying temperatures (21 °C, 40 °C, 60 °C, 80 °C), NaCl concentrations (2 м, 4 м, 6 м), and pH levels (2, 3, 4, 7) for a fixed duration of 5 minutes. As anticipated, the bending curvature increased with higher stimulus intensities across all three types of stimuli. For instance, with temperature variations, the bilayer quickly bends towards the PNIPAM/SA/NC side, and the most significant curvature change, from 0.00 to 0.89 $mm^{-1}$, occurs between 21 °C and 40 °C, aligning with the LCST of PNIPAM (32 °C). [22] Additionally, when exposed to 80 °C, the average curvature increases to 1.6 $mm^{-1}$ (Figure 3c). Regarding the ionic response of the bilayers, a more pronounced change in curvature, from 0.67 to 1.01 $mm^{-1}$, is observed between 4 м and 6 м NaCl, indicating that a higher salt concentration accelerates the shape-morphing (Figure 3c). However, in acidic solutions, the bilayer bends towards the PAA/SA/NC side, resulting in negative average curvature values. In response to pH, the most significant alteration in curvature, from -0.18 to -0.39 mm$^{-1}$, happens between pH 4 and pH 3, the closest range to the pKa (4.3) of PAA. [52] When exposed to pH 2, the bilayer reaches a higher average curvature of -0.5 mm$^{-1}$ (Figure 3c).

The shape-morphing results of the bilayers are attributed to the material properties of each hydrogel layer. For example, the bending curvature of the bilayer immersed in 6 M NaCl is lower than that of the bilayer at 80 °C, despite both stimuli inducing similar area changes in the PNIPAM/SA/NC hydrogel (Figure 2g). This difference in curvature may be attributed to osmotic processes occurring in the PAA/SA/NC hydrogel layer and/or salt aggregation within the bilayer's structure at high NaCl concentrations, which could restrict the movement of the PNIPAM/SA/NC layer. Moreover, while the PAA/SA/NC hydrogel alone demonstrates greater responsivity than the PNIPAM/SA/NC hydrogel, evidenced by its higher values of shrinking and area change upon exposure to pH (Figure 2), the bilayer system exhibits the lowest curvature when exposed to pH compared to the other stimuli. This can be attributed to the effect of CaCl$_2$ crosslinking and the difference in stiffness between the two hydrogels. While the shrinking and area-change experiments did not involve CaCl$_2$ crosslinking of the samples, the bilayer structures were crosslinked with 1 wt% of CaCl$_2$ to facilitate strong interfacial bonding between the layers with the SA network. However, it was observed that even this low concentration of CaCl$_2$ significantly reduces (by a factor of 11) the swelling capacity of the PAA/SA/NC hydrogel (Figure S3, Supporting Information). Therefore, the shrinking properties of the PAA/SA/NC hydrogel in the bilayer system are limited due to the CaCl$_2$ crosslinking. Furthermore, the mechanical properties of the PAA/SA/NC hydrogel are comparatively lower than those of the PNIPAM/SA/NC hydrogel. Even in its shrunk state, after 24 hours in pH 2, the PAA/SA/NC hydrogel exhibits lower stiffness (178 kPa) than the swollen PNIPAM/SA/NC hydrogel (229 kPa) (Figure 2e,h). Consequently, when the bilayer is immersed in pH 2, the PNIPAM/SA/NC layer is stiffer than the PAA/SA/NC layer. Thus, although the PAA/SA/NC layer contracts due to shrinking, its actuation is constrained by the stiffer PNIPAM/SA/NC layer on top of it.

Understanding the bilayer's mechanism of bidirectional bending with varied curvatures in response to different types and intensities of stimuli enables programmability of the actuator's behavior for complex scenarios, including sequential environmental changes. This capability was explored by subjecting the same bilayer sample to a sequence of three stimuli: first temperature (60 °C), then pH (2), and finally NaCl (6 M), each for 5 minutes. Initially, under the temperature stimulus, the PNIPAM/SA/NC layer contracts, causing a closing movement. When subsequently immersed in an acidic solution at room temperature (21 °C), the PNIPAM/SA/NC layer becomes passive, and the PAA/SA/NC layer activates, resulting in an opening movement. Finally, in the salt solution, the PAA/SA/NC layer becomes passive again, while the PNIPAM/SA/NC layer reactivates, leading to the restoration of the closing movement. This sequence highlights the bilayer system's ability to function as both a mechanical sensor and actuator, sensing its surrounding environment and undergoing programmed actuation due to its multi-stimuli responsiveness.

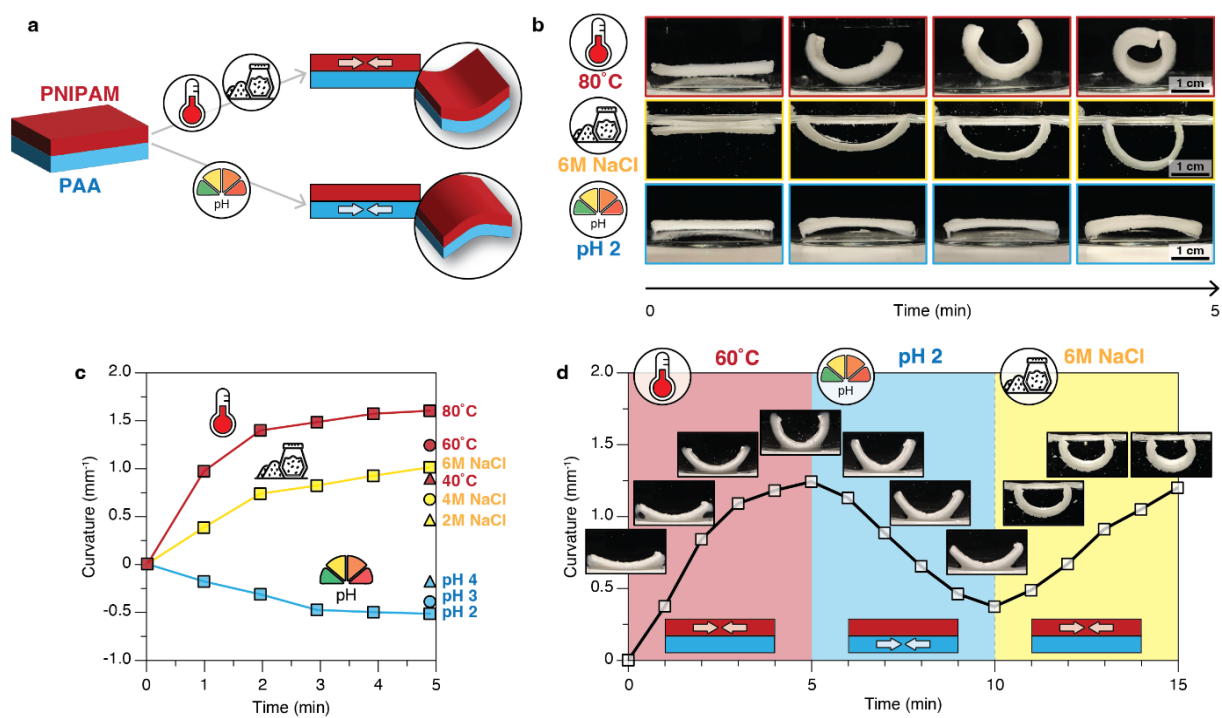

**Figure 3.** Bilayer system composed of PNIPAM/SA/NC and PAA/SA/NC hydrogels exhibits bending actuation in response to diverse stimulus scenarios. (A) Schematics of bilayer actuation resulting from the shrinkage of each layer triggered by external stimuli (temperature, and salt concentration to PNIPAM/SA/NC, and pH to PAA/SA/NC). (B) Bidirectional bending of the bilayers observed during a 5-minute exposure to temperature (80 °C), NaCl (6 м), and pH (2). (C) Alteration in curvature of the bilayers in response to varying stimuli and their intensities. (D) Shape-morphing of a bilayer during sequential exposure to stimuli over time: 0 to 5 minutes at 60 °C, 5 to 10 minutes at pH 2, and 10 to 15 minutes at NaCl concentration of 6 м.

### 2.4. Materials Logic

The possibility to modulate the actuation of the developed bilayers by programming which layer is activated or remains passive during exposure to various stimuli scenarios, including combinations of stimuli, enables the programmability of the bilayers to conduct logic operations. In Boolean logic-based systems, binary logic gates operate with two logic levels ("0" and "1") in both input and output states.[61] Here, input signals are binary: "0" is assigned when the stimulus is OFF, and "1" when the stimulus is ON. By combining these binary signals corresponding to temperature, NaCl, and pH stimuli, it is possible to generate eight distinct input signals, one for each logic operation, based on the presence and combination of stimuli **(Figure 4a)**. Each signal comprises three digits: the first denotes the temperature stimulus, the second denotes NaCl, and the third denotes pH. For instance, the logic operation signal (1 0 1) indicates that temperature and pH are present inputs. Additionally, the logic operations can be represented by Venn diagrams. Each segment of the Venn diagram corresponds to a unique combination of inputs and signifies whether the stimulus is present (colored: red for temperature, yellow for NaCl, and blue for pH) or absent (white) (Figure 4b). To establish the operational logic of the developed system, combinations of stimuli are applied by exposing the bilayers to two or three stimuli simultaneously. While we utilized a binary code for the inputs of the logic operations, the outputs cannot be simply binarized, as they vary based on the resulting shape of the bilayers in response to the applied stimuli combination. Unlike conventional mechanical logic gates that generate discrete digital signals, chemical logic gates rely on responsive materials and exhibit analog signals, dependent on the stimuli.[62]

Hence, the outputs of the logic operations are assessed based on the curvature of the bilayers following the fixed time of 5 minutes of actuation under each of the eight combinations of stimuli. A different bilayer sample, not previously exposed to any stimulus, was used for each scenario of stimuli combination. The operation (1 1 0) exhibits the highest bilayer curvature value of 1.75 mm$^{-1}$ (Figure 4c). This result arises from the combined effects of the contributing stimuli: temperature (80 °C) and NaCl (6 м), which amplify the shrinking of the PNIPAM/SA/NC layer, leading to a more pronounced bending actuation. Conversely, the (0 0 0) operation, where no stimulus is applied, results in close to zero shape transformation. The second lowest positive curvature value of 0.30 mm$^{-1}$ is observed for the (0 1 1) operation, resulting from the combination of NaCl (6 м) and pH (2), two competing stimuli that induce moderate curvature in both layers (Figure 4c). Similarly, the operation (1 0 1) exhibits a lower curvature actuation of 0.45 mm$^{-1}$, also attributed to the presence of competing stimuli (temperature and pH), which trigger the actuation of both layers (Figure 4c). In this scenario, the curvature surpasses that of operation (0 1 1) because temperature induces a greater actuation of the PNIPAM/SA/NC layer than NaCl (Figure 3c). Furthermore, for both operations (0 1 1) and (1 0 1), the bending curvature remains positive despite the presence of pH stimulus, which typically induces negative bending, indicating that the actuation force of the PNIPAM/SA/NC layer is stronger than that of the PAA/SA/NC layer. The operation (0 0 1) presents the most negative curvature value of -0.52 mm$^{-1}$, attributed to the single input of pH (2) activating the PAA/SA/NC layer, resulting in negative bending (Figure 4c). On the other hand, the operations (1 0 0) and (0 1 0) exhibit high curvature values of 1.60 mm$^{-1}$ and 1.01 mm$^{-1}$, respectively, due to the robust actuation of the PNIPAM/SA/NC layer in response to temperature (80 °C) and NaCl (6 M), respectively (Figure 4c). The operation (1 1 1), which combines all the three stimuli (80 °C, 6 M NaCl, pH 2) displays a moderate curvature value of 0.87 mm$^{-1}$ due to the simultaneous shrinking of both PNIPAM/SA/NC and PAA/SA/NC layers (Figure 4c).

Following Boolean logic principles, NOT, YES, OR, and AND gates were constructed, resulting in eight distinct chemical logic gates, each corresponding to a different stimuli combination (Figure 4d). Stimuli-responsive materials effectively function as YES gates, operating on an ON-OFF principle as a switch mechanism. However, to construct Boolean logic-based systems, materials need to react to multiple inputs to form AND, OR, and other logic gates. Moreover, these materials must react independently, without subsequent or coordinated responses.[62] An OR gate generates an output upon sensing either of two inputs. While numerous multi-stimuli-responsive material systems exist, OR gate necessitates that either stimulus is adequate to induce a response and that they provoke the same response.[62] AND gates demand two inputs to yield a single output. In responsive materials, AND gates enhance specificity, thereby increasing efficiency.[17] Here, a NOT gate is formed by the absence of any stimulus, operation (0 0 0). YES gates are created by exposing the bilayer to one stimulus at a time, either temperature (1 0 0), NaCl (0 1 0), or pH (0 0 1) (Figure 4d). An OR gate is formed by subjecting the bilayer to both temperature and NaCl at the same time (1 1 0), resulting in intense bending actuation, due to the combination of two contribute stimuli (Figure 4d). In contrast, AND gates are based on the interaction of competing stimuli. Two AND gates are created by exposing the bilayer to pH and temperature (1 0 1), or to pH and NaCl (0 1 1), causing both layers to shrink simultaneously (Figure 4d). Additionally, a more intricate logic operation is conducted by combining an OR gate with an AND gate. When the bilayer is exposed to all stimuli at the same time (1 1 1), both layers become active. The temperature and NaCl stimuli contribute to an OR gate, which feeds its output as one of the inputs, along with pH (a competing stimulus), to an AND gate, resulting in moderate bending actuation (Figure 4d). Therefore, the materials logic developed here relies on chemical logic gates yielding mechanical actuation outputs. These analog outputs, derived from the curvature of the bilayers, offer advantages over the digital outputs of conventional logic materials. One benefit lies in sensing subtle nuances in data.[62]

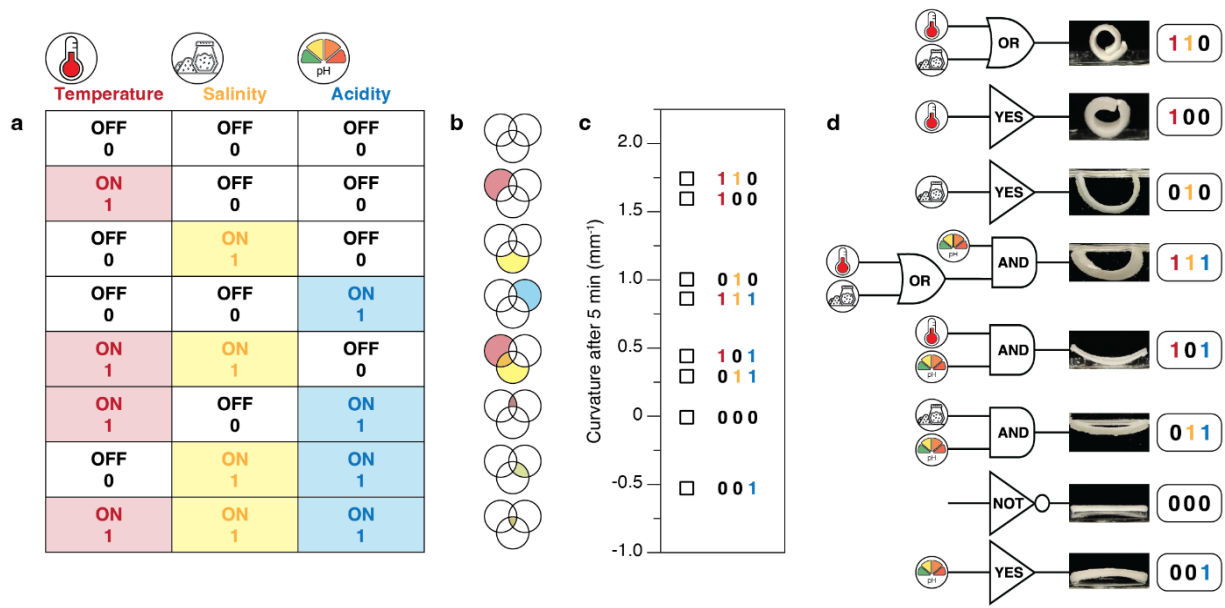

**Figure 4.** Logic operations performed by bilayer structures in response to combinations of stimuli (temperature, salt concentration, and pH). (A) Input table showing binary signals denoted by the presence (ON = 1) or absence (OFF = 0) of stimuli. (B) Corresponding Venn diagrams illustrating the presence of stimuli and overlapping regions. (C) Bending curvature observed for each of the eight logic operations. (D) Representation of logic gates (YES, NOT, OR, AND) with stimuli inputs and resulting shape-morphing outputs.

## 3. Conclusions

We have designed and fabricated actuators with embodied logic using responsive hydrogels (PAA and PNIPAM) reinforced with high concentration of nanocellulose (14 wt% CNC and 1 wt% CNF) through 4D printing. By adjusting the stimuli, we determined the physical properties of the multi-responsive composites, including swelling capacity, stiffness, and shape-morphing ability, necessary to characterize their response under different environmental conditions. This enabled both temporal and spatial programmability of the bidirectional bending actuation in bilayer structures triggered by temperature, NaCl, and pH variations. The combination of these stimuli generated various responsiveness scenarios, which we used to demonstrate NOT, YES, OR, and AND logic gates, accomplishing eight different logic operations. The analog outputs (a range of intermediate curvatures) from these operations highlight the potential of the developed material as a mechanical sensor with a simple optical readout. Further exploration of the anisotropic properties provided by the alignment of the nanocellulose particles could enable extra logic operations. Additionally, integrating more degrees of programmability into the materials by tailoring the crosslinking density in different regions of the structure could induce localized properties, further enhancing the versatility of the actuators. Therefore, the reported programmable multi-responsive hydrogels provide a novel approach to designing logic material systems that can be leveraged for autonomous soft robotics, biomedical diagnostic, drug delivery, smart agriculture, and environmental monitoring and remediation.

## 4. Experimental Section

*Materials*

For the TEMPO-mediated oxidation: 2,2,6,6-tetramethyl-1-piperidinyloxyl (TEMPO), sodium bromide (NaBr) 99%, and sodium hypochlorite solutions (NaClO) (12-14% chlorine) were purchased from VWR International (Belgium). Sodium hydroxide (NaOH) 99% was acquired from Carl Roth GmbH + Co. KG (Germany). For the preparation of the precursor gels: lyophilized cellulose nanocrystals (CNCs) from acid hydrolysis of eucalyptus pulp were acquired from CelluForce (Canada), and cellulose fibers, derived from bleached wood pulp, were used for the production of cellulose nanofibers (CNFs). The monomers *N*-isopropylacrylamide (NIPAM) 97%, and acrylic acid (AA) 99%, as well as the crosslinking agent *N,N'*-methylenebis(acrylamide) (MBA) 99%, sodium alginate (SA), calcium chloride ($CaCl_2$), sodium chloride (NaCl), and hydrochloric acid (HCl) were acquired from Sigma-Aldrich. The photoinitiator lithium phenyl-2,4,6-trimethylbenzoylposphinate (LAP) 99% was purchased from Apollo Scientific (England).

*Fabrication of Cellulose Nanofibers (CNF)*

Never-dried cellulose fibers were TEMPO oxidized following a well-established protocol from Saito and Isogai [63], which is extensively described elsewhere. [48] The oxidized cellulose fibers were separated from the solution, by using metal and cotton fabric filters, and were thoroughly washed with distilled water for 5 days, until the conductivity was similar to that of distilled water. The oxidized cellulose fibers were dispersed in deionized water to a concentration of 2 wt% and grounded using a Supermass Colloider (MKZA10- 20 J CE Masuko Sangyo, Japan) to obtain a suspension of cellulose nanofibers (CNFs). The energy applied to the grinding process was 10 kWh per kg of cellulose.

*Preparation of Nanocellulose-based Inks*

The dispersion of the cellulose nanoparticles (14 wt% CNC, and 1 wt% CNF) and the other ingredients (75.44 wt% of deionized water, 1.00 wt% of SA, 8.25 wt% of NIPAM, 0.06 wt% of MBA, and 0.25 wt% of LAP) was achieved by mechanical mixing of the inks either using a speedmixer (SpeedMixer DAC 150.1 FVZ) at speeds 1000, 1500, 2000, and 2350 rpm for 1 minute each and repeating this program for 3 times, or using an overhead mechanical stirrer (BDC3030, Caframo) at 1300 rpm for 15 minutes with a metallic rod and a 4 cm cross-shaped impeller. The same procedure has been adopted for the AA ink, with the only alteration being the substitution of NIPAM with AA (7.5 wt%) and an increase in the SA content to 2 wt%. The nanocellulose-based inks were stored in the fridge (4 °C) for one night. Before printing, the gel was filled into plastic syringe cartridges and centrifuged (centrifuge ROTINA-380 Hettich, or 5804 Eppendorf) for 4 minutes at 3500 rpm to remove air bubbles.

*Rheology of Inks*

The rheological behavior of the developed inks was characterized using a rotational rheometer (MCR 302, Anton Paar), and using non-crosslinked hydrogels, without the presence of the photoinitiator. Measurements were carried out with a parallel plates geometry, with a diameter of 50 mm and a spacing of 0.5 mm at a constant temperature of 20 °C. The flow viscosity was obtained by varying the rotational shear rate from 0.001 to 1000 $s^{-1}$ with a logarithmic sweep. With the amplitude sweeps, the elastic shear modulus (G') and viscous modulus (G") were measured using logarithmic oscillatory intervals at a frequency of 1 Hz (deformation variations from 0.01 to 1000%). The apparent yield stress was defined as the shear stress when the storage and loss moduli intersect, i.e., the gel point.

*3D Printing*

Nanocellulose-based hydrogels were 3D printed using the direct ink writing (DIW) technique with equipment from EnvisionTEC (Bioplotter Manufacturing Series) and 3D Systems (Allevi 2). The hydrogels were loaded into plastic cartridges and extruded through uniform steel nozzles (J.A. Crawford co.) at room temperature (21 °C), utilizing compressed air at pressures ranging from 1.1 to 1.7 bar and speeds between 8.0 and 11.5 mm $s^{-1}$. The extrusion needles were 25.0 mm long and had a non-tapered geometry with diameters of 0.7 mm for the AA ink and 0.4 mm for the NIPAM ink.

*Crosslinking*

Following the 3D printing process, the hydrogel structures underwent crosslinking through two sequential steps: (1) photopolymerization of the PNIPAM or PAA network, by using ultraviolet (UV) light irradiation in a nitrogen ($N_2$) atmosphere to prevent oxygen inhibition of the reaction, the structures were placed within a chamber equipped with UV lamps (50 W power, 365 nm wavelength, Everbeam), positioned 5 cm away from the lamps, for 3 to 10 minutes, depending on the dimensions and thickness of the printed structure; (2) crosslinking the SA network, by immersing the printed and UV-cured structures in an aqueous solution containing either 1 wt% or 5 wt% calcium chloride (CaCl$_2$) for 24 hours, this immersion facilitates ionic crosslinking between the alginate molecules and Ca$^{2+}$ ions.

*Swelling Characterization*

Time-dependent swelling tests were conducted to quantify the maximal swelling capability of the hydrogels. Printed cubic samples (1x1x1 cm) with 100% infill and non-crosslinked with CaCl$_2$ were used. For the swelling tests, initially, the samples were completely dried in an oven at 60 °C for 4 hours. Then, they were weighed and submerged in distilled water at room temperature for varying durations, ranging from 5 minutes to 5 days. Following each specified period, the samples were removed from the water, placed on a paper towel to remove excess water from the surface, and then weighed again in their swollen state. A minimum of 3 samples of the same material were measured for each swelling period. The swelling percentage, also known as equilibrium moisture content (EMC), of the material from fully dried hydrogel samples was calculated (Equation 1).

$$Swelling\ (\%) = \frac{W_s - W_d}{W_d} * 100 \qquad (1)$$

Where *Ws* is the weight of the swollen sample, and *Wd* is the weight of the dry sample.

*Shrinking Characterization*

To determine the shrinking percentage of the hydrogels in response to different stimuli, fully swollen PNIPAM/SA/NC hydrogel samples (after 4 days immersed in room temperature DI water) were submerged in DI water at 40 °C, 60 °C, and 80 °C, as well as in solutions with varying NaCl concentrations (2 м, 4 м, 6 м) for durations ranging from 2 minutes to 3 days. Similarly, fully swollen PAA/SA/NC hydrogels were immersed in acid solutions with pH 2, 3, and 4, also for various controlled periods of time. At least 3 samples of the same material and stimuli conditions were measured for each shrinking period. The shrinking capacity of the hydrogels was assessed in terms of the loss of water from the samples compared to their swollen weight (Equation 2).

$$Shrinking\ (\%) = \frac{W_s - W_c}{W_s} * 100 \qquad (2)$$

Where *Ws* is the weight of the swollen sample, and *Wc* is the weight of the shrunk sample.

*Area Change Characterization*

In order to investigate the change in size of the PNIPAM/SA/NC and PAA/SA/NC hydrogels during swelling and shrinking tests, top view pictures of each sample were taken for 10 hours of experiment. The top view area of the cubic samples was measured using the image analysis software ImageJ and the percentage of area change after 24h of swelling was calculated (Equation 3).

$$Area\ change\ in\ swelling\ (\%) = \frac{A - A_0}{A_0} * 100 \quad (3)$$

Where $A$ is the area of the swollen sample, and $A_0$ is the area of the dried sample.

Similarly, it was calculated the percentage of area change after 24h of shrinking (Equation 4).

$$Area\ change\ in\ shrinking\ (\%) = \frac{A_0 - A}{A_0} * 100 \quad (4)$$

Where $A_0$ is the area of the fully swollen sample, and $A$ is the area of the shrunk sample.

*Mechanical Characterization*

To assess the mechanical properties of the hydrogels, cubic samples (1x1x1 cm) with 100% infill were printed and subjected to compression testing using a universal testing machine (ElectroPuls E3000, INSTRON) equipped with a 250 N load cell or (Zwick Roell - Z005 Universal Testing System) with a 100 N load cell, operating at a rate of 1 mm min$^{-1}$. Additionally, samples of the hydrogel matrices without nanocellulose reinforcement were manually cast and cut, as printing these solutions was not feasible due to their low viscosity. Prior to compression testing, samples were immersed in a solution of 5 wt% of CaCl$_2$ for 24h of crosslinking, and then they were placed in DI water for 4 days until they reached their fully swollen state. Paper towels were used to gently remove excess water from the surfaces of the specimens. For the stimuli responsiveness tests, the samples were exposed to variations in temperature, NaCl concentration, and pH for 24 hours prior to compression testing. At least 3 specimens were examined for each type of hydrogel and each stimulus parameter. The stiffness of the materials was determined by the elastic modulus, or Young's modulus, which was calculated by analyzing the slope of the elastic region at the initial stage of the stress vs. strain curve.

*Bilayer Structure Fabrication*

A layer of PNIPAM/SA/NC hydrogel was printed on top of a layer of PAA/SA/NC hydrogel, having both layers 100% infill and filament orientation of 0°. The multi-material-bilayers were crosslinked with 1 wt% CaCl$_2$ solution for 24 hours.

*Shape-morphing Characterization*

The bilayer structures were submerged in various solutions to expose them to different intensities of stimuli: temperature (40 °C, 60 °C, 80 °C), NaCl concentration (2 м, 4 м, 6 м), and pH levels (2, 3, 4); individually as well as in combinations, for a duration of 5 minutes each. Front-view photos of the samples were captured, and the average curvature of each sample was calculated using the Kappa plugin [64] in the image processing software Fiji (ImageJ). [65]

**Supporting Information**

Supporting Information is available from the Wiley Online Library or from the authors.

**Acknowledgements**

The authors thank Mathilde Champeau for helpful discussions, and Chelsea Fox for helping with the operation of the universal testing machine. B.A. and S.T. acknowledge the São Paulo Research Foundation (FAPESP) for the financial support under grants #2021/00380-4, #2021/10037-5, #2022/10706-7, and #2023/04970-6. T.M. acknowledges support from the Swiss National Science Foundation (P500PT_203197/1). C.D. acknowledges support from the MURI ARO W911NF-22-2-0109.

**Author Contributions**

B.A., G.S., T.M., G.N., S.T., and C.D. designed the study. B.A. conducted the experiments, processed the data, and wrote the manuscript with input from all coauthors. All authors reviewed and commented on the manuscript.

**References**

[1] Y. Dong, A. N. Ramey-Ward, K. Salaita, *Advanced Materials* **2021**, *33*.
[2] X. Fan, J. Y. Chung, Y. X. Lim, Z. Li, X. J. Loh, *ACS Appl Mater Interfaces* **2016**, *8*, 33351.
[3] P. Lavrador, M. R. Esteves, V. M. Gaspar, J. F. Mano, *Adv Funct Mater* **2021**, *31*.
[4] F. G. Woodhouse, J. Dunkel, *Nat Commun* **2017**, *8*.
[5] B. Mazzolai, A. Mondini, E. Del Dottore, L. Margheri, F. Carpi, K. Suzumori, M. Cianchetti, T. Speck, S. K. Smoukov, I. Burgert, T. Keplinger, G. D. F. Siqueira, F. Vanneste, O. Goury, C. Duriez, T. Nanayakkara, B. Vanderborght, J. Brancart, S. Terryn, S. I. Rich, R. Liu, K. Fukuda, T. Someya, M. Calisti, C. Laschi, W. Sun, G. Wang, L. Wen, R. Baines, S. K. Patiballa, R. Kramer-Bottiglio, D. Rus, P. Fischer, F. C. Simmel, A. Lendlein, *Multifunctional Materials* **2022**, *5*.
[6] X. Xia, C. M. Spadaccini, J. R. Greer, *Nat Rev Mater* **2022**, *7*, 683.
[7] P. Jiao, J. Mueller, J. R. Raney, X. (Rayne) Zheng, A. H. Alavi, *Nat Commun* **2023**, *14*.
[8] S. B. Choi, *Front Mater* **2014**, *1*.
[9] C. El Helou, P. R. Buskohl, C. E. Tabor, R. L. Harne, *Nat Commun* **2021**, *12*.
[10] H. Yasuda, P. R. Buskohl, A. Gillman, T. D. Murphey, S. Stepney, R. A. Vaia, J. R. Raney, *Nature* **2021**, *598*, 39.
[11] A. Pal, M. Sitti, *Proc Natl Acad Sci U S A* **2023**, *120*.

[12]  Y. Song, R. M. Panas, S. Chizari, L. A. Shaw, J. A. Jackson, J. B. Hopkins, A. J. Pascall, *Nat Commun* **2019**, *10*.
[13]  Y. Jiang, L. M. Korpas, J. R. Raney, *Nat Commun* **2019**, *10*.
[14]  B. Treml, A. Gillman, P. Buskohl, R. Vaia, *Proc Natl Acad Sci U S A* **2018**, *115*, 6916.
[15]  T. Mei, Z. Meng, K. Zhao, C. Q. Chen, *Nat Commun* **2021**, *12*.
[16]  H. Komatsu, S. Matsumoto, S. ichi Tamaru, K. Kaneko, M. Ikeda, I. Hamachi, *J Am Chem Soc* **2009**, *131*, 5580.
[17]  B. A. Badeau, M. P. Comerford, C. K. Arakawa, J. A. Shadish, C. A. Deforest, *Nat Chem* **2018**, *10*, 251.
[18]  X. Zhang, S. Soh, *Advanced Materials* **2017**, *29*.
[19]  A. S. Hoffman, *Adv Drug Deliv Rev* **2012**, *64*, 18.
[20]  Lh. Yahia, *Journal of Biomedical Sciencies* **2015**, *04*.
[21]  E. M. Ahmed, *J Adv Res* **2015**, *6*, 105.
[22]  L. Tang, L. Wang, X. Yang, Y. Feng, Y. Li, W. Feng, *Prog Mater Sci* **2021**, *115*.
[23]  M. J. Ansari, R. R. Rajendran, S. Mohanto, U. Agarwal, K. Panda, K. Dhotre, R. Manne, A. Deepak, A. Zafar, M. Yasir, S. Pramanik, *Gels* **2022**, *8*.
[24]  X. Xu, Y. Liu, W. Fu, M. Yao, Z. Ding, J. Xuan, D. Li, S. Wang, Y. Xia, M. Cao, *Polymers (Basel)* **2020**, *12*.
[25]  Y. Guan, Y. Zhang, *Soft Matter* **2011**, *7*, 6375.
[26]  S. Dutta, D. Cohn, *J Mater Chem B* **2017**, *5*, 9514.
[27]  L. S. Lim, N. A. Rosli, I. Ahmad, A. Mat Lazim, M. C. I. Mohd Amin, *Nanomaterials* **2017**, *7*.
[28]  P. Cao, L. Tao, J. Gong, T. Wang, Q. Wang, J. Ju, Y. Zhang, *ACS Appl Polym Mater* **2021**, *3*, 6167.
[29]  J. Lai, X. Ye, J. Liu, C. Wang, J. Li, X. Wang, M. Ma, M. Wang, *Mater Des* **2021**, *205*.
[30]  A. Kirillova, R. Maxson, G. Stoychev, C. T. Gomillion, L. Ionov, *Advanced Materials* **2017**, *29*.
[31]  M. L. Oyen, *International Materials Reviews* **2014**, *59*, 44.
[32]  L. E. Beckett, J. T. Lewis, T. K. Tonge, L. S. T. J. Korley, *ACS Biomater Sci Eng* **2020**, *6*, 5453.
[33]  S. E. Bakarich, R. Gorkin, M. In Het Panhuis, G. M. Spinks, *Macromol Rapid Commun* **2015**, *36*, 1211.
[34]  K. J. De France, E. D. Cranston, T. Hoare, *ACS Appl Polym Mater* **2020**, *2*, 1016.
[35]  G. Siqueira, D. Kokkinis, R. Libanori, M. K. Hausmann, A. S. Gladman, A. Neels, P. Tingaut, T. Zimmermann, J. A. Lewis, A. R. Studart, *Adv Funct Mater* **2017**, *27*.
[36]  N. Peng, D. Huang, C. Gong, Y. Wang, J. Zhou, C. Chang, *ACS Nano* **2020**, *14*, 16169.
[37]  L. Y. Ee, S. F. Yau Li, *Nanoscale Adv* **2021**, *3*, 1167.
[38]  O. Fourmann, M. K. Hausmann, A. Neels, M. Schubert, G. Nyström, T. Zimmermann, G. Siqueira, *Carbohydr Polym* **2021**, *259*.
[39]  M. K. Hausmann, P. A. Rühs, G. Siqueira, J. Läuger, R. Libanori, T. Zimmermann, A. R. Studart, *ACS Nano* **2018**, *12*, 6926.
[40]  A. Sydney Gladman, E. A. Matsumoto, R. G. Nuzzo, L. Mahadevan, J. A. Lewis, *Nat Mater* **2016**, *15*, 413.
[41]  C. Gauss, K. L. Pickering, L. P. Muthe, *Composites Part C: Open Access* **2021**, *4*.
[42]  S. Tibbits, *Proc. of the 34th Annual Conf. of the Association for Computer Aided Design for Architecture*, Los Angeles, October **2014**.
[43]  J. Choi, O. C. Kwon, W. Jo, H. J. Lee, M. W. Moon, *3D Print Addit Manuf* **2015**, *2*, 159.
[44]  M. Champeau, D. A. Heinze, T. N. Viana, E. R. de Souza, A. C. Chinellato, S. Titotto, *Adv Funct Mater* **2020**, *30*.

[45]  W. Liu, L. Geng, J. Wu, A. Huang, X. Peng, *Compos Sci Technol* **2022**, *225*.
[46]  Y. C. Huang, Q. P. Cheng, U. S. Jeng, S. H. Hsu, *ACS Appl Mater Interfaces* **2023**, *15*, 5798.
[47]  M. C. Mulakkal, R. S. Trask, V. P. Ting, A. M. Seddon, *Mater Des* **2018**, *160*, 108.
[48]  R. Weishaupt, G. Siqueira, M. Schubert, P. Tingaut, K. Maniura-Weber, T. Zimmermann, L. Thöny-Meyer, G. Faccio, J. Ihssen, *Biomacromolecules* **2015**, *16*, 3640.
[49]  K. Matsumoto, N. Sakikawa, T. Miyata, *Nat Commun* **2018**, *9*.
[50]  Y. Chen, W. Xu, W. Liu, G. Zeng, *J Mater Res* **2015**, *30*, 1797.
[51]  X. Yan, Y. Chu, B. Liu, G. Ru, Y. Di, J. Feng, *Physical Chemistry Chemical Physics* **2020**, *22*, 12644.
[52]  T. Swift, L. Swanson, M. Geoghegan, S. Rimmer, *Soft Matter* **2016**, *12*, 2542.
[53]  M. K. Hausmann, G. Siqueira, R. Libanori, D. Kokkinis, A. Neels, T. Zimmermann, A. R. Studart, *Adv Funct Mater* **2020**, *30*.
[54]  D. Buenger, F. Topuz, J. Groll, *Prog Polym Sci* **2012**, *37*, 1678.
[55]  T. Gwan Park, A. S. Hoffman, *Sodium Chloride-Induced Phase Transition in Nonionic Poly(N-Isopropylacrylamide) Gel*, **1993**.
[56]  I. Burgert, P. Fratzl, *Philosophical Transactions of the Royal Society A: Mathematical, Physical and Engineering Sciences* **2009**, *367*, 1541.
[57]  D. Correa, S. Poppinga, M. D. Mylo, A. S. Westermeier, B. Bruchmann, A. Menges, T. Speck, *Philosophical Transactions of the Royal Society A: Mathematical, Physical and Engineering Sciences* **2020**, *378*.
[58]  P. Fratzl, R. Elbaum, I. Burgert, *Faraday Discuss* **2008**, *139*, 275.
[59]  R. M. Erb, J. S. Sander, R. Grisch, A. R. Studart, *Nat Commun* **2013**, *4*.
[60]  J. W. Boley, W. M. Van Rees, C. Lissandrello, M. N. Horenstein, R. L. Truby, A. Kotikian, J. A. Lewis, L. Mahadevan, *Proc Natl Acad Sci U S A* **2019**, *116*, 20856.
[61]  S. B. Jo, J. Kang, J. H. Cho, *Advanced Science* **2021**, *8*.
[62]  E. M. Bressler, S. Adams, R. Liu, Y. L. Colson, W. W. Wong, M. W. Grinstaff, *Clin Transl Med* **2023**, *13*.
[63]  T. Saito, A. Isogai, *Biomacromolecules* **2004**, *5*, 1983.
[64]  H. Mary, G. J. Brouhard, (Preprint) bioRxiv: 10.1101/852772, submitted: Nov **2019.**
[65]  J. Schindelin, I. Arganda-Carreras, E. Frise, V. Kaynig, M. Longair, T. Pietzsch, S. Preibisch, C. Rueden, S. Saalfeld, B. Schmid, J. Y. Tinevez, D. J. White, V. Hartenstein, K. Eliceiri, P. Tomancak, A. Cardona, *Nat Methods* **2012**, *9*, 676.

# Supporting Information

**Programmable Multi-responsive Nanocellulose-based Hydrogels with Embodied Logic**

*Beatriz Arsuffi[1,2,3], Gilberto Siqueira[2] \*, Gustav Nyström[2], Silvia Titotto[3] \*,*
*Tommaso Magrini[1,4] \*, Chiara Daraio[1] \**

1 - Division of Engineering and Applied Science, California Institute of Technology, Pasadena, CA 91125, USA.
2 - Empa, Swiss Federal Laboratories for Materials Science and Technology, Cellulose and Wood Materials Laboratory, 8600, Dübendorf, Switzerland.
3 - Center of Engineering, Modelling and Applied Social Sciences, Federal University of ABC, Santo André, SP, Brazil.
4 - Department of Mechanical Engineering, Eindhoven University of Technology, 5600MB Eindhoven, The Netherlands.

\* Email: gilberto.siqueira@empa.ch, silvia.titotto@ufabc.edu.br, t.magrini@tue.nl, daraio@caltech.edu

Legend

Supporting Information S1: Rheological properties of the inks

Supporting Information S2: Crosslinking effect

Supporting Information S3: Shrinking behavior

Supporting Information S4: Area change

Supporting Information S5: Effects of nanocellulose reinforcement and double-network system on the mechanical properties of the 3D printed composites

Supporting Information S6: Mechanical properties

Supporting Information S7: Shape-morphing

Supporting Information Refences

**Supporting Information S1: Rheological properties of the inks**

The rheological behavior of the nanocellulose-based inks was characterized to evaluate their suitability for 3D printing using the direct ink writing (DIW) technique. Previous works have shown that to achieve good printability through DIW, three main rheological properties are required: (1) G' (storage modulus, associated with the material's elasticity) should exceed G" (loss modulus, related to viscosity); (2) G" should be higher than a few kPa; and (3) the yield stress should fall within the range of 100 Pa. [1,2] Here, rheology tests were conducted on inks containing NIPAM, SA, and nanocellulose (15 wt% CNC, or 14 wt% CNC and 1 wt% CNF). The positive effect of the nanocellulose reinforcement on the viscosity of the inks was verified. At shear rates exceeding 1 $s^{-1}$, the matrix ink (without nanocellulose) exhibits low viscosity of 1.12 Pa.s, indicating its propensity to flow directly through the print nozzle at low pressures and lack the ability to present structural integrity post-extrusion (Figure S1a). At higher shear rates, shear-thinning behavior is observed only for the inks with CNC and CNF, attributed to the alignment of the cellulose particles (Figure S1a). Due to their pseudoplasticity, these inks demonstrate viscosities ranging from 7.5 Pa.s (ink with 15 wt% CNC) to 14.0 Pa.s (ink with 14 wt% CNC and 1 wt% CNF) at a shear rate of 100 $s^{-1}$, indicating their suitability for DIW.

To evaluate the viscoelastic properties of the inks, oscillatory rheological measurements were conducted at a frequency of 1 Hz (Figure S1b). These measurements reveal that both cellulose inks demonstrate elastic behavior at low shear rates (G' > G"), affirming their stability and elastic response as a solid. When G' equals G", a well-defined dynamic flow is observed, indicating the transition between solid and liquid behavior, as G" surpasses G', and both values decrease by several orders of magnitude. The dynamic yield stress ranges from 26 to 399 Pa for inks containing 15 wt% CNC, and 14 wt% CNC plus 1 wt% CNF, respectively, underscoring the notable impact of adding 1 wt% CNF on the rheological properties of the inks (Figure S1b).

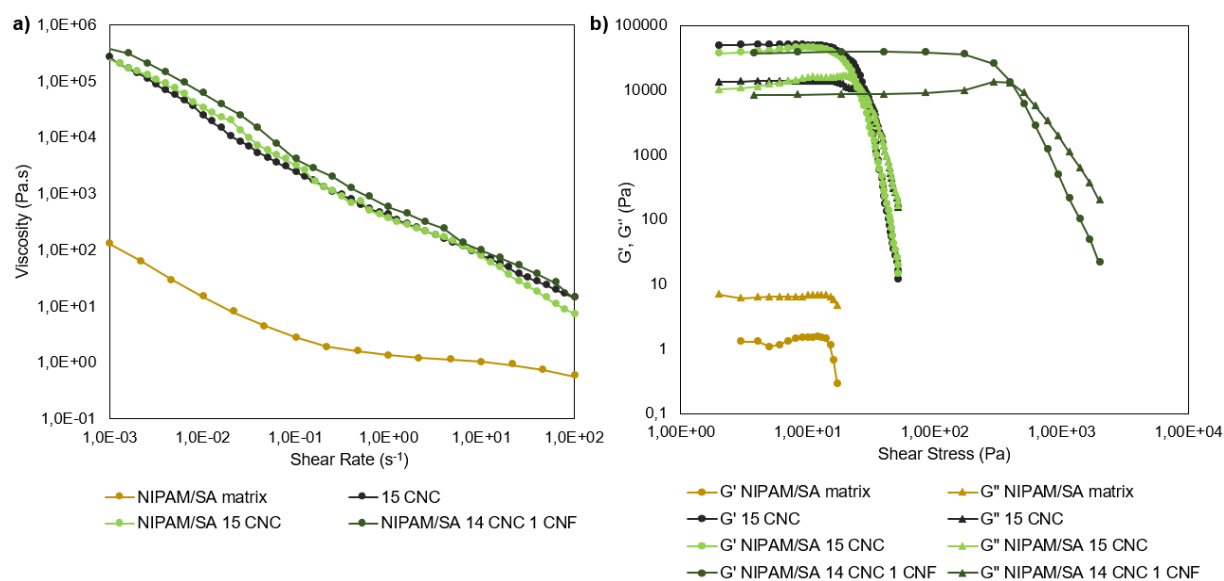

**Figure S1.** Rheological properties of the NIPAM/SA inks with varied nanocellulose loading (15 wt% CNC and 14 wt% CNC plus 1 wt% CNF). **a)** Shear sweep (steady-shear measurements). **b)** Amplitude sweep (oscillatory measurements).

**Supporting Information S2: Crosslinking effect**

The crosslinking process of the 3D printed hydrogel structures occurred in two steps. Initially, immediately after printing, as the ink was still unpolymerized, the structures lacked shape stability and were deformable when handled (Figure S2). The first crosslinking step involved the photopolymerization of the NIPAM or AA monomer through UV irradiation in a nitrogen ($N_2$) atmosphere for 5 to 20 min, depending on the thickness of the structure, to prevent oxygen inhibition of the polymerization reaction. The second stage of crosslinking involved immersing the structures in a calcium chloride ($CaCl_2$) solution (1 wt% or 5 wt%) for 24 hours, facilitating the ionic crosslinking of SA with $Ca^{2+}$ ions. Following this procedure, the structures demonstrated both shape stability and flexibility, capable of reverting to their original shape after manual deformation (Figure S2). Furthermore, they did not dissolve over a six-month period when submerged in water or when subjected to various environmental stimuli, including changes in temperature, salt concentration, and pH.

The crosslinking with $CaCl_2$ solution significantly affects the swelling capacity and mechanical properties of the PNIPAM/SA/NC and PAA/SA/NC hydrogels. As the $CaCl_2$ concentration increases, the swelling capacity of both hydrogels decreases due to the intensified crosslinking of the SA network, which restricts water uptake (Figure S3). Conversely, as the $CaCl_2$ concentration increases, the stiffness of the hydrogels also increases because of the higher crosslinking density within the SA network (Figure S4).

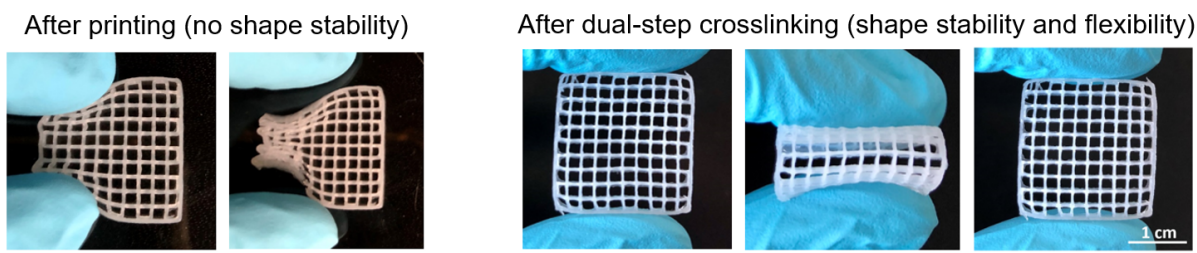

**Figure S2.** Effect of crosslinking on 3D printed structures of PNIPAM/SA/NC hydrogel.

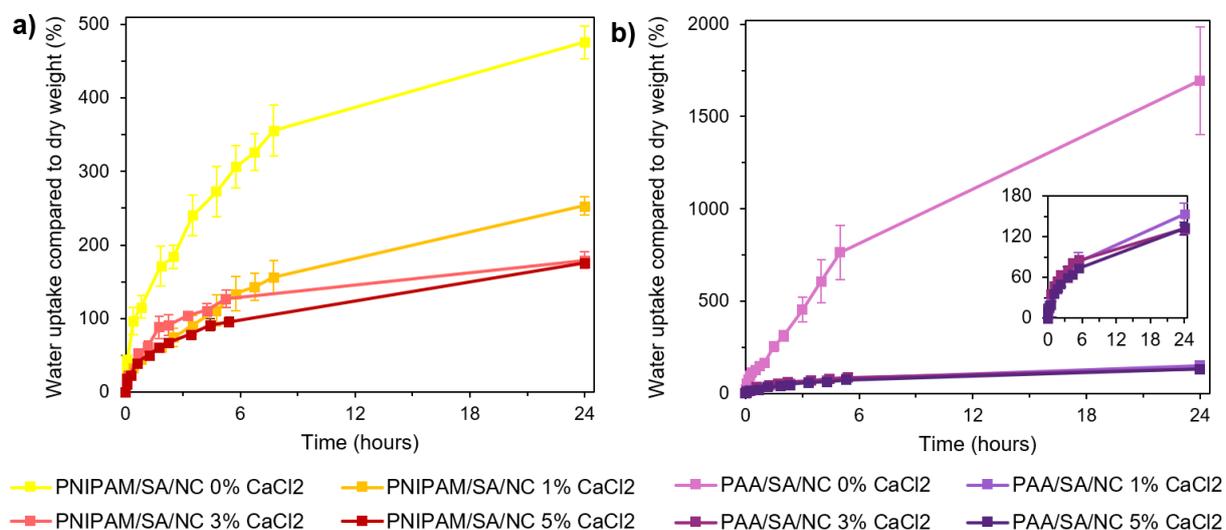

**Figure S3.** Effect of the $CaCl_2$ crosslinking on the swelling capacity of the cast: **a)** PNIPAM/SA/NC and **b)** PAA/SA/NC hydrogels in 24 hours. Error bars show standard deviation (n = 3).

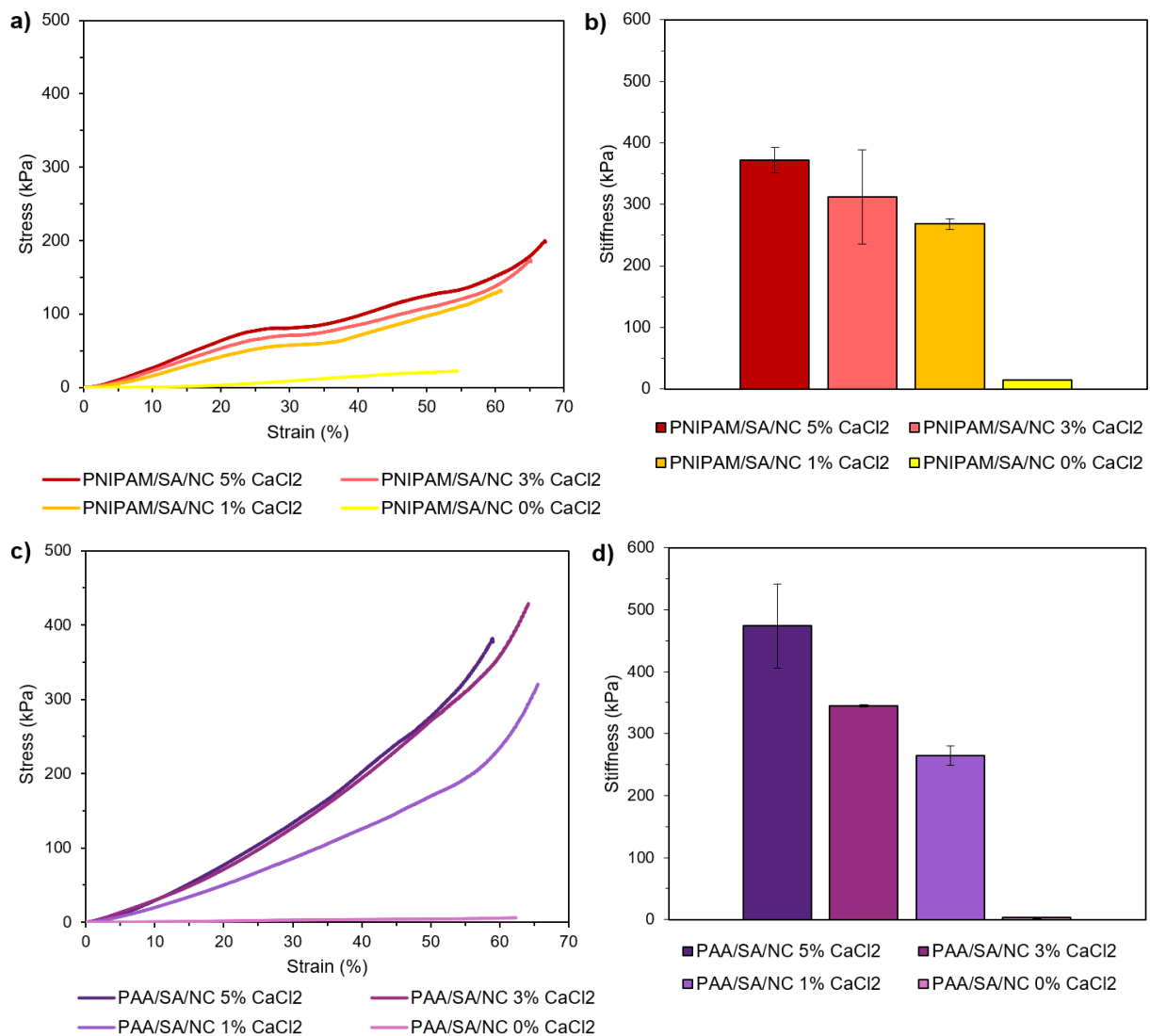

**Figure S4.** Effect of the CaCl$_2$ crosslinking on the mechanical properties of cast hydrogels. **a)** Average stress vs. strain curves of PNIPAM/SA/NC. **b)** Young's modulus. **c)** Average stress vs. strain curves of PAA/SA/NC. **d)** Young's modulus. Error bars show standard deviation (n = 3).

**Supporting Information S3: Shrinking behavior**

Given that PNIPAM and PAA are stimuli-responsive hydrogels, we investigated the shrinking behavior of the developed materials when exposed to environmental stimuli using 3D printed cubic samples (1x1x1 cm). The shrinking kinetics of the PNIPAM/SA/NC hydrogel was investigated for 3 days in response to temperatures above PNIPAM's lower critical solution temperature (LCST = 32 °C). As expected, the PNIPAM/SA/NC hydrogel demonstrates accelerated shrinking behavior with increasing temperature. For example, within 15 minutes of shrinking, the sample at 40 °C exhibits a water loss of 8% compared to its fully swollen state, while the samples at 60 °C and 80 °C experience water losses of 29% and 38%, respectively

(Figure S5b). However, when comparing the extent of water loss after 1 day of shrinking, the difference related to temperature is less than 4% and the shrinking rates stabilize (Figure S5a). This suggests that regardless of the temperature applied, provided it is higher than 32 °C, the material reaches approximately the same maximum shrinking value (around 59% of water loss) after a longer period of exposure to heat. This loss of water is attributed to the transition of PNIPAM's polymer chains from a hydrophilic to a hydrophobic state at its LCST.[3] Moreover, when temperature (80 °C) and NaCl (6 M) stimuli are combined, the presence of anions in the surrounding media additionally reduces the LCST of PNIPAM[4], resulting in a faster shrinking behavior. For instance, after only 5 minutes of shrinking the PNIPAM/SA/NC hydrogel presents a water loss value higher than 40% (Figure S5b). Concerning the shrinking of the PNIPAM/SA/NC hydrogel in response to the NaCl stimulus alone, the salt concentration plays a significant role in determining the water loss values after prolonged exposure (after 1 day) (Figure S5c). Furthermore, the shifting shrinking rate regarding the NaCl concentration over time indicates that salt aggregation on the samples could influence the water loss results (Figure S5c,d). In accordance with the swelling outcomes, the PAA/SA/NC hydrogel exhibits greater shrinkage compared to the PNIPAM/SA/NC hydrogel (Figure S5e). When submerged in acid solutions with a pH below PAA's pKa (4.3), the PAA/SA/NC hydrogel promptly shrinks and achieves nearly maximal shrinkage rates in less than 1 hour, highlighting its rapid shrinking kinetics (Figure S5f). For instance, after 30 minutes in the acid solutions, the sample in pH 2 exhibits the highest water loss value with 73%, followed by the sample in pH 3 with 61%, and then by the one in pH 4 with 7%, indicating a corresponding increase in water loss with the decrease in the pH value (Figure S5f). After 3 days of immersion in the acid solutions, only the sample in pH 4 shows a higher water loss value (22%) compared to the value within the first 30 minutes of the experiment (Figure S5e). Meanwhile, the samples in pH 3 and pH 2 exhibit nearly identical shrinking rates to those observed within the first 30 minutes of the experiment, with 77% and 80% of water loss, respectively, demonstrating significant differences in shrinking behavior when the pH is reduced to 3 compared to pH 4, which might be very close to PAA's pKa, limiting the full pH responsivity of the PAA/SA/NC hydrogel.

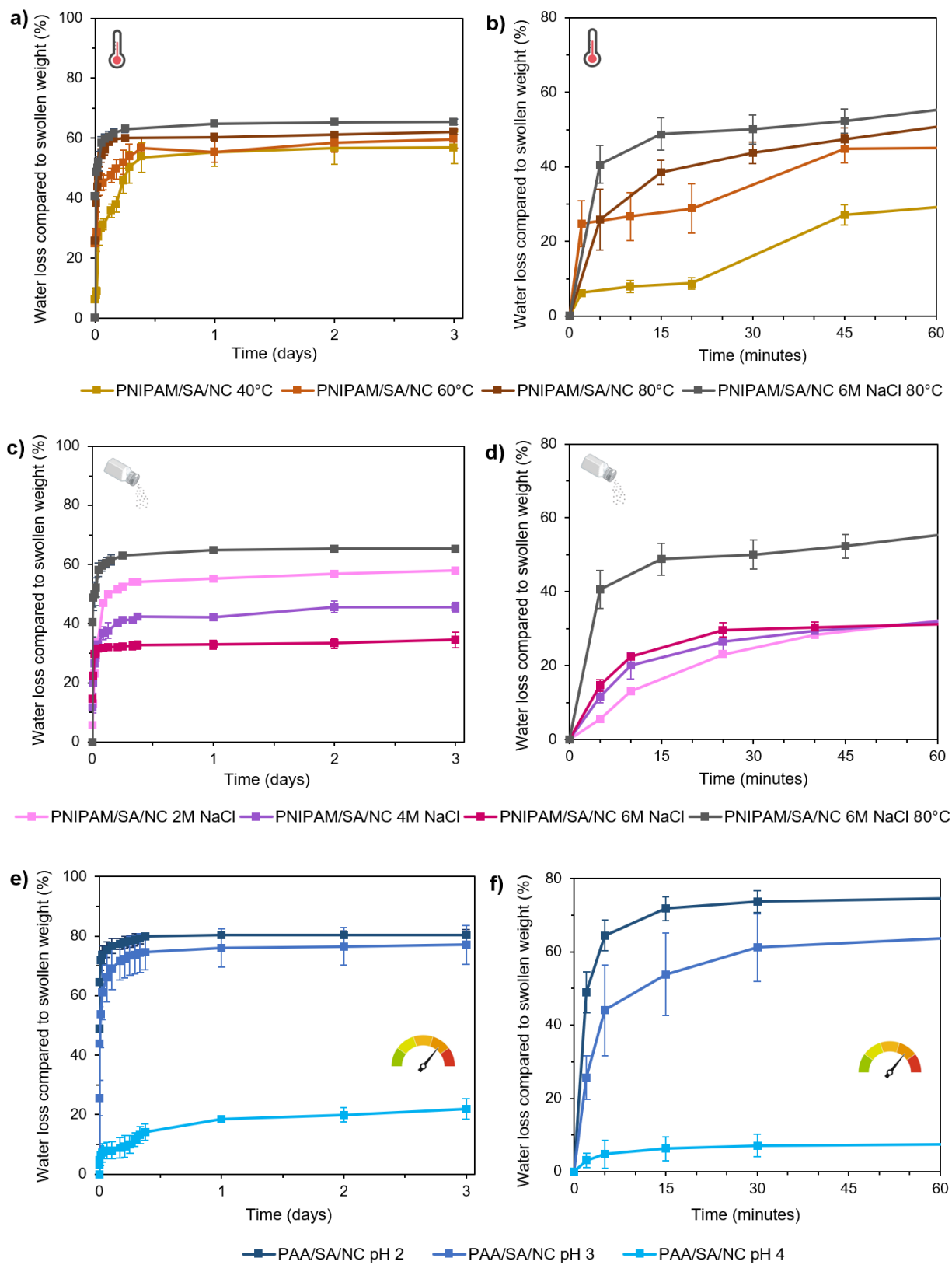

**Figure S5.** Effect of stimuli on the shrinking behavior of 3D printed hydrogels. **a)** and **b)** PNIPAM/SA/NC under temperature. **c)** and **d)** PNIPAM/SA/NC exposed to NaCl solutions. **e)** and **f)** PAA/SA/NC in acid solutions. Error bars show standard deviation (n = 3).

**Supporting Information S4: Area change**

To assess the shape oscillation caused by the swelling and shrinking mechanisms of the developed hydrogels, the change in area of the samples was investigated (Figure S6). As expected, the increase in area of the samples during swelling mirrors the behavior observed in the swelling curves for both PNIPAM/SA/NC and PAA/SA/NC hydrogels (Figure S7). This increase occurs as the sample absorbs water, causing its polymeric network to expand and consequently leading to an increase in volume. The PAA/SA/NC hydrogel demonstrates a greater swelling capacity compared to the PNIPAM/SA/NC hydrogel. Consequently, it exhibits a significantly larger increase in size after 24 hours of swelling, expanding by 1228% compared to its dried state. While the PNIPAM/SA/NC hydrogel only enlarges by 383% relative to its initial dried area after absorbing water for the same duration (Figure S7).

Likewise, consistent with its shrinking response, the PNIPAM/SA/NC hydrogel contracts in size when subjected to heat (Figure S6b). For example, after 24 hours of exposure to DI water at 40 °C, 60 °C, and 80 °C, the samples show reductions in area of 42%, 44%, and 48%, respectively. This aligns with the observation that, after 24 hours of shrinking at various temperatures, the PNIPAM/SA/NC samples demonstrate similar levels of water loss (Figure S8a). Conversely, the alteration in area of the PNIPAM/SA/NC hydrogel when submerged in NaCl solutions does not correspond directly to its shrinking behavior. The samples exposed to higher concentrations of NaCl exhibit greater reduction in size. For instance, the hydrogels immersed in 2 M, 4 M, and 6 M NaCl solutions for 24 hours demonstrate 32%, 35%, and 49% of area change, respectively (Figure S8b). This indicates that the water loss values of PNIPAM/SA/NC hydrogel immersed in NaCl solutions were indeed affected by the salt aggregation on the samples. Furthermore, as anticipated, the samples of PNIPAM/SA/NC hydrogel that display higher contraction due to more pronounced shrinking are those exposed to the combination of temperature (80 °C) and NaCl (6 M) with 59% reduction in size (Figure S8d). Regarding the area change of the PAA/SA/NC hydrogels during shrinking, the samples immersed in pH 3 and 2 exhibit the fastest and most significant contraction, reducing their areas by more than 61% and 62%, respectively (Figure S8c). Meanwhile, the samples immersed in pH 4 demonstrate the lowest change in area compared to all tested samples, with an average of 13% change in area, aligning with their lower shrinking capacity (Figure S8c).

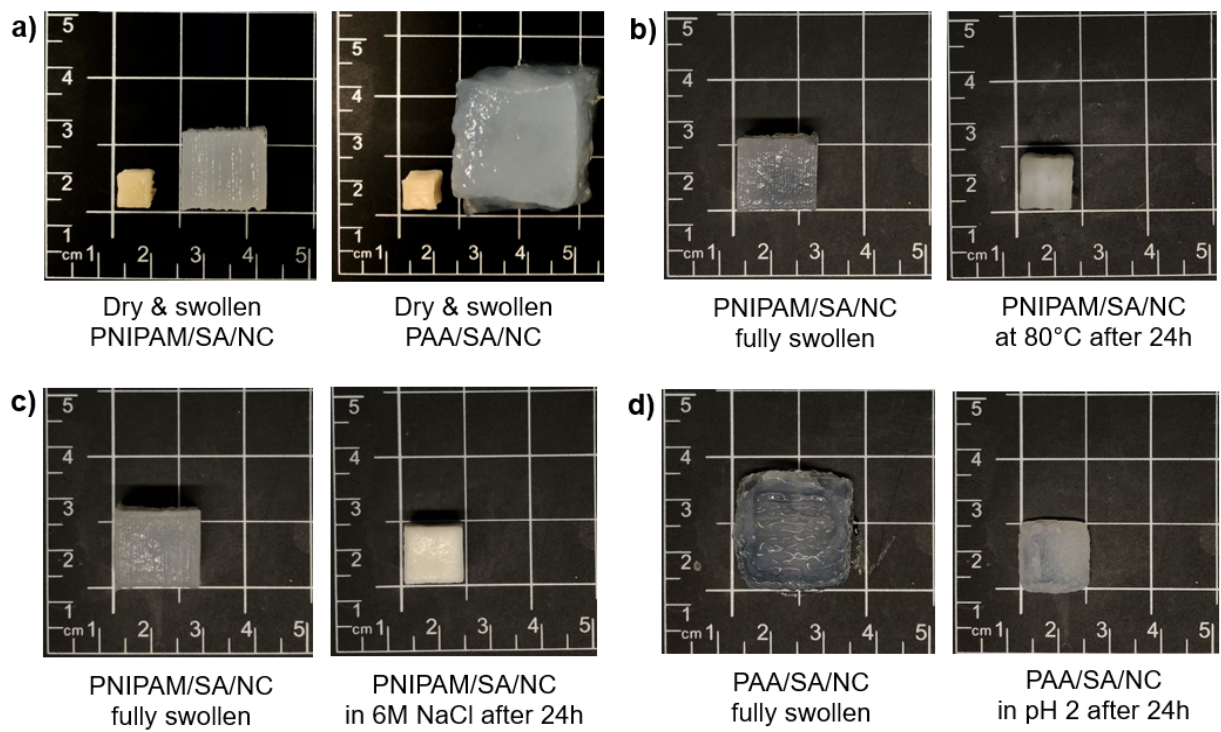

**Figure S6.** Comparison of the size of the 3D printed hydrogels due to swelling and shrinking. **a)** PNIPAM/SA/NC and PAA/SA/NC hydrogels in their dry and swollen (after 24 h in DI water) states. **b)** PNIPAM/SA/NC hydrogel before and after shrinking at 80 °C for 24h. **c)** PNIPAM/SA/NC hydrogel before and after shrinking in 6 M NaCl for 24h. **d)** PAA/SA/NC hydrogel before and after shrinking in pH 2 for 24h.

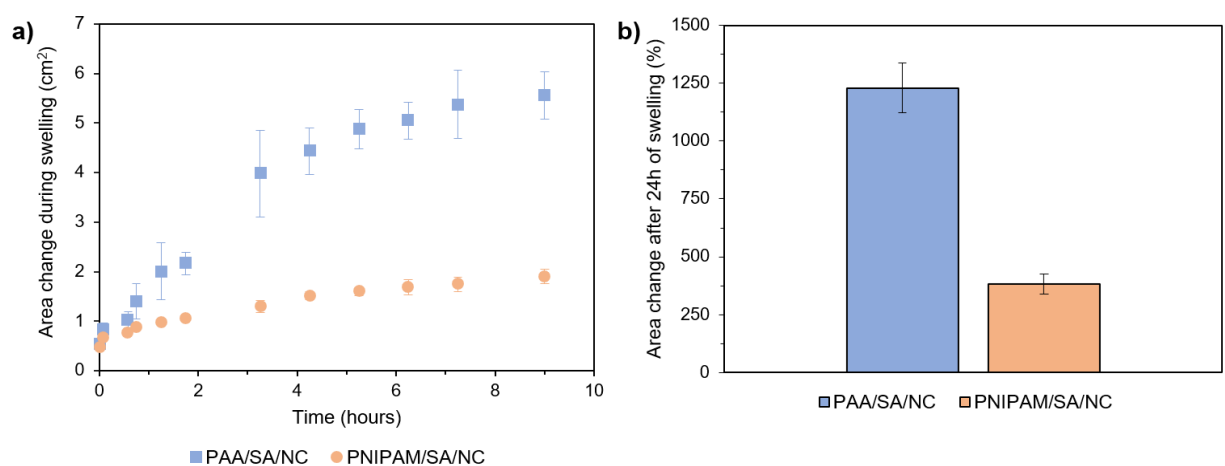

**Figure S7.** Change in area during swelling of the PNIPAM/SA/NC and PAA/SA/NC hydrogels. **a)** Area oscillation over time of swelling. **b)** Area change after 24 hours of swelling. Error bars show standard deviation (n = 3).

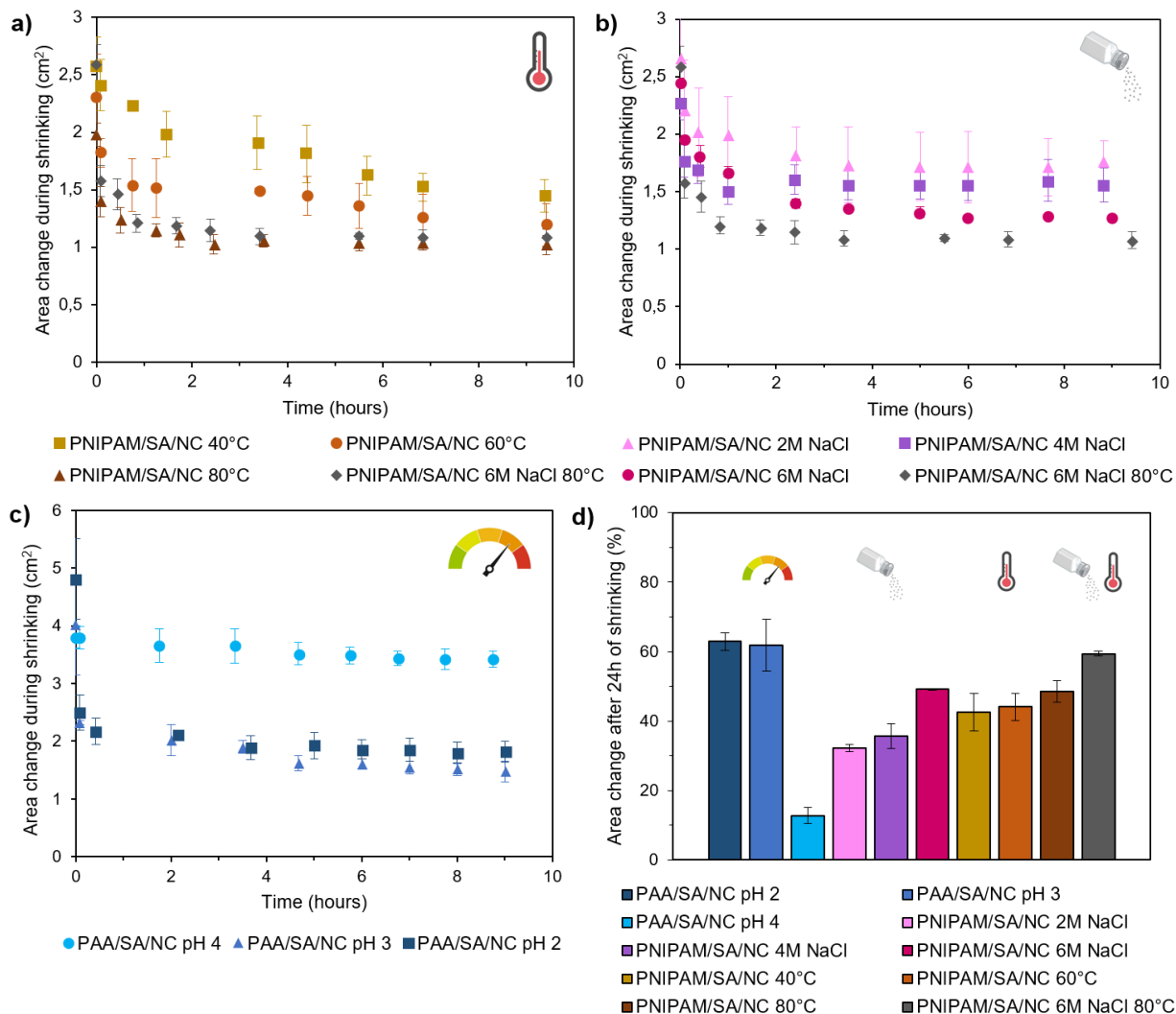

**Figure S8.** Change in area during shrinking of the PNIPAM/SA/NC and PAA/SA/NC hydrogels in response to: **a)** temperature, **b)** NaCl, and **c)** pH. **d)** Area change after 24 hours of shrinking. Error bars show standard deviation (n = 3).

**Supporting Information S5: Effects of nanocellulose reinforcement and double-network system on the mechanical properties of the 3D printed composites**

To investigate the effect of nanocellulose (both CNC and CNF) reinforcement on the stiffness of the hydrogels, 3D printed samples (1x1x1 cm) of PNIPAM/SA/NC hydrogels, as well as cast samples of PNIPAM/SA matrix without nanocellulose content were tested in compression (Zwick Roell - Z005 Universal Testing System, 100 N load cell, rate of 1 mm min$^{-1}$). While the hydrogel matrix alone exhibits Young's modulus of 5.26 kPa, the 3D printed sample of the formulation with 15 wt% CNC tested in the transverse direction shows significantly higher Young's modulus, averaging 186.70 kPa (Figure S9). This reinforcing effect is particularly evident when comparing the stiffness of the formulation that includes 1 wt% of CNF along with the 14 wt% CNC, showing a Young's modulus of 252.94 kPa, demonstrating the beneficial influence of adding 1 wt% CNF into the inks (Figure S9).

Moreover, the double-network structure of the hydrogels developed in this work resulted in significantly higher stiffness compared to those reported in the literature for nanocellulose-reinforced PNIPAM printed hydrogels. In the work of Fourmann et al. [5], the unreinforced single-network PNIPAM hydrogel matrix exhibits Young's modulus of 70 Pa, whereas the double-network PNIPAM/SA matrix developed in this work presents a modulus 75 times higher (5.26 kPa) (Figure S9). Additionally, the 3D printed samples of PNIPAM/SA/NC containing 14 wt% CNC and 1 wt% CNF presents Young's modulus more than 13 times higher than that reported by Fourmann et al. [5] for samples with 20 wt% CNC (Figure S9). This highlights the positive impact of incorporating 1 wt% of sodium alginate (SA) in the formulation of the developed hydrogel, enhancing its stiffness through the establishment of a double-network matrix with PNIPAM, along with the addition of 1 wt% CNF.

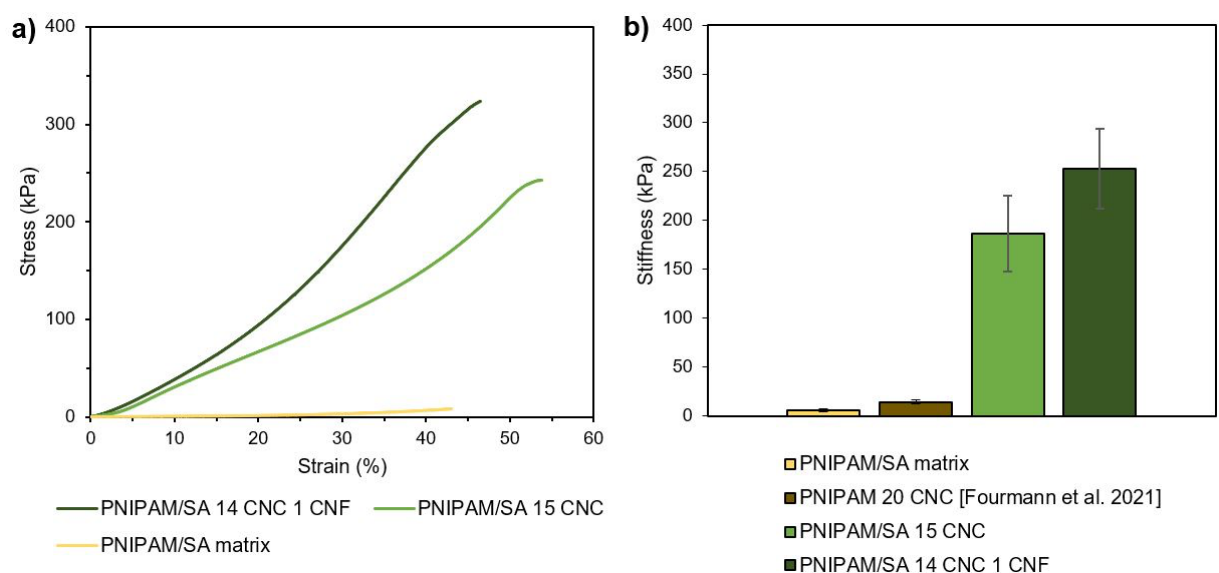

**Figure S9.** Effect of nanocellulose reinforcement on the mechanical properties of 3D printed PNIPAM/SA/NC hydrogels and cast PNIPAM/SA matrix. **a)** Average stress vs. strain curves. **b)** Young's modulus. Error bars show standard deviation (n = 7).

**Supporting Information S6: Mechanical properties**

To assess the impact of external stimuli on the mechanical properties of the responsive hydrogels, 3D printed samples previously subjected to variations in temperature, NaCl concentration, and pH for 24 hours were tested in compression (ElectroPuls E3000, INSTRON, 250 N load cell, rate of 1 mm min$^{-1}$). Due to PNIPAM's thermoresponsive nature, the

PNIPAM/SA/NC hydrogel exhibits increased stiffness with rising temperatures. For instance, while the PNIPAM/SA/NC hydrogel at 20 °C displays a Young's modulus of 228.64 kPa, the same hydrogel at 80 °C demonstrates a modulus of 323.54 kPa (Figure S10a,b). By testing samples at 20 °C, 40 °C, 60 °C, and 80 °C, it is observed that the highest gain on Young's modulus (a difference of 46 kPa) is shown in samples tested at 40 °C compared to those tested at 20 °C, attributable to the volume phase transition of PNIPAM occurring at 32 °C. Samples tested at 60 °C exhibit a 26 kPa increase in Young's modulus compared to those tested at 40 °C. Similarly, samples tested at 80 °C demonstrate a 23 kPa difference compared to those tested at 60 °C (Figure S10a,b). Likewise, the shrinking effect in the stiffness of the PNIPAM/SA/NC hydrogel is even more remarkable for the samples immersed in NaCl solutions (Figure S10c,d). The PNIPAM/SA/NC hydrogel previously immersed in 6 M NaCl exhibits a Young's modulus of 1782.69 kPa, nearly eight times higher than that of the same hydrogel not exposed to NaCl solution. As expected, the stiffness of the hydrogels increases progressively with the rise in salt concentration. The most significant increase in Young's modulus is observed in samples previously exposed to 2 M NaCl compared to those in DI water, showing a difference of 1225.84 kPa (Figure S10c,d). Upon comparison of samples immersed in 2 M, 4 M, and 6 M NaCl, it is evident that the average stiffness increases by 100 kPa with each incremental rise of 2 M in NaCl concentration (Figure S10c,d). Given the high NaCl concentration, the increase in stiffness may also be influenced by salt aggregation within the printed structures in addition to water loss. When both temperature (80 °C) and NaCl (6 M) stimuli are combined, the presence of anions in the surrounding media further reduces the LCST of PNIPAM. [4] Consequently, the PNIPAM/SA/NC hydrogel exhibits a Young's modulus of 3373.84 kPa due to a more intense shrinking process (Figure S10c,d). This corresponds to an increase of the Young's modulus by a factor of 14.7 when compared to the modulus of the sample in DI water. Finally, the stiffness of the PAA/SA/NC hydrogels increases with the decrease in pH of the solutions, as PAA undergoes shrinking when immersed in pH values below its pKa (around 4.3). [6] The PAA/SA/NC hydrogels tested at pH 7, pH 4, and pH 3 demonstrate very similar Young's modulus values: 123.08 kPa, 123.32 kPa, and 134.65 kPa, respectively (Figure S10e,f). However, a more pronounced increase in stiffness is observed for samples at pH 2, exhibiting a Young's modulus of 178.55 kPa (Figure S10e,f).

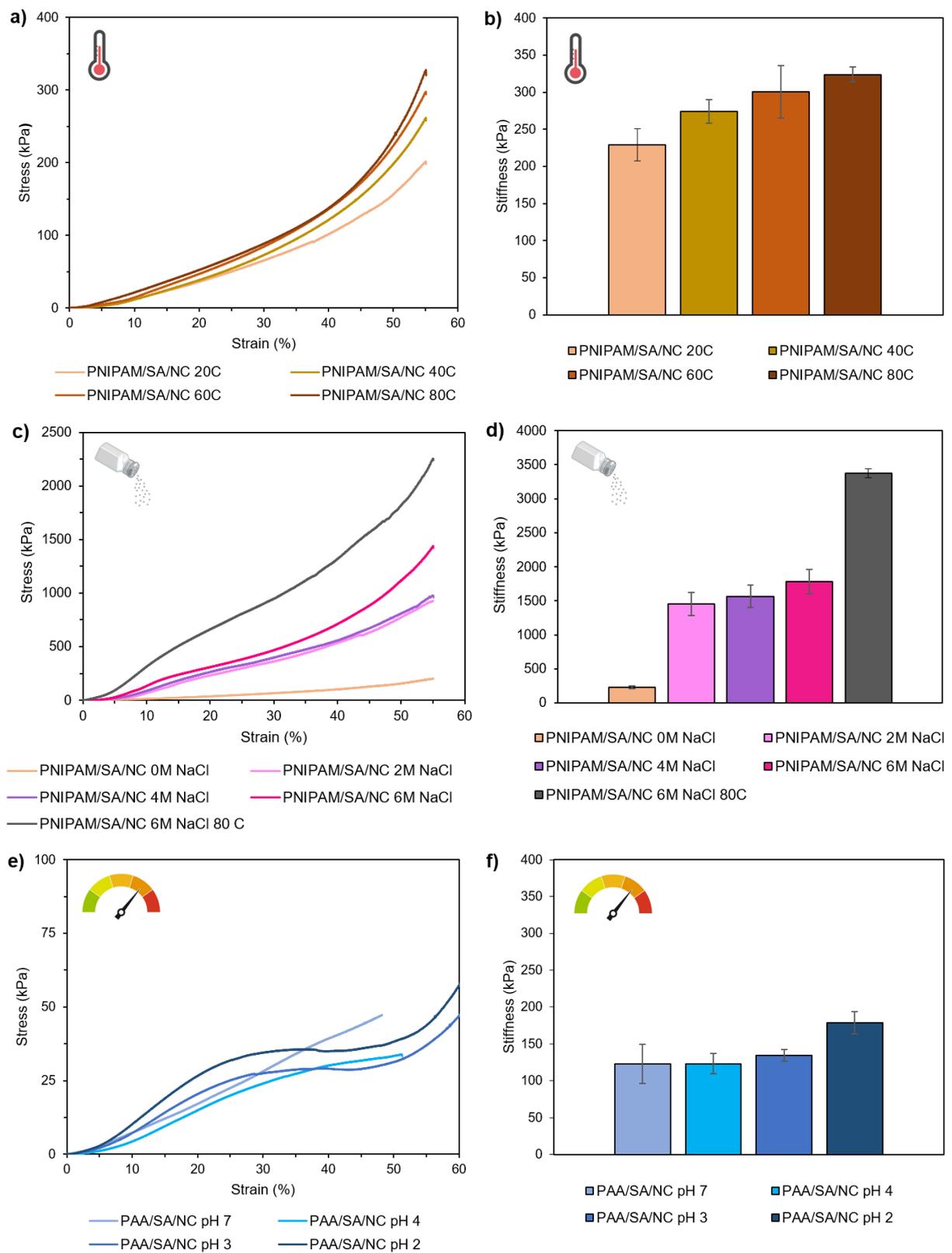

**Figure S10.** Effect of stimuli on the mechanical behavior of 3D printed hydrogels. **a)** Average stress vs. strain curves of PNIPAM/SA/NC under temperature. **b)** Young's modulus. **c)** Average stress vs. strain curves of PNIPAM/SA/NC in NaCl solutions. **d)** Young's modulus. **e)** Average stress vs.

strain curves of PAA/SA/NC in acid solutions. **f)** Young's modulus. Error bars show standard deviation (n = 3).

**Supporting Information S7: Shape-morphing**

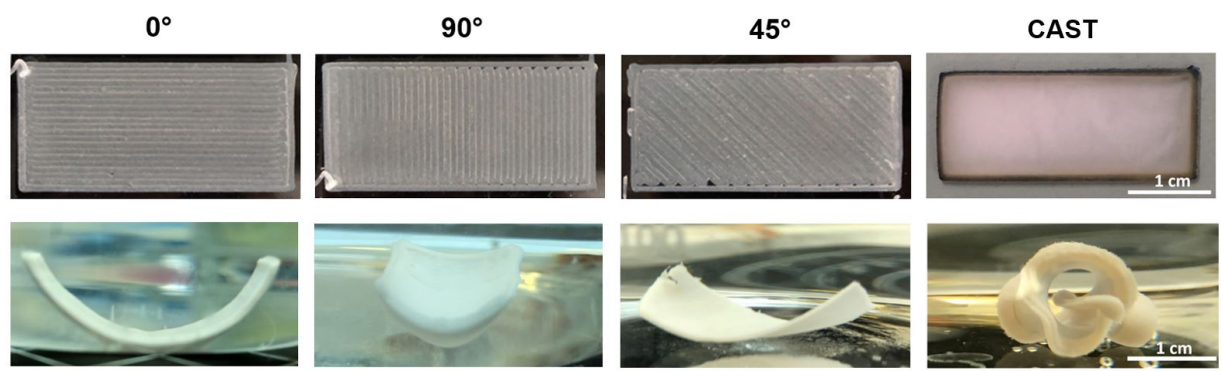

**Figure S11.** Anisotropic shape-morphing of 3D printed PNIPAM/SA/NC hydrogel bilayers (30x10x0.8 mm) compared to a cast sample. The first horizontal series of images exhibits monolayers printed with a larger needle diameter (0.8 mm) to visually demonstrate the variation in filament extrusion (0°, 45°, and 90°). The subsequent series of images illustrates the shape-morphing of bilayers printed with a smaller needle diameter (0.4 mm) in response to temperature (60 °C) after 5 minutes.

**Supporting Information References**

[1] G. Siqueira, D. Kokkinis, R. Libanori, M. K. Hausmann, A. S. Gladman, A. Neels, P. Tingaut, T. Zimmermann, J. A. Lewis, A. R. Studart, *Adv Funct Mater* **2017**, *27*.

[2] M. K. Hausmann, P. A. Rühs, G. Siqueira, J. Läuger, R. Libanori, T. Zimmermann, A. R. Studart, *ACS Nano* **2018**, *12*, 6926.

[3] K. Matsumoto, N. Sakikawa, T. Miyata, *Nat Commun* **2018**, *9*.

[4] T. Gwan Park, A. S. Hoffman, *Sodium Chloride-Induced Phase Transition in Nonionic Poly(N-Isopropylacrylamide) Gel*, **1993**.

[5] O. Fourmann, M. K. Hausmann, A. Neels, M. Schubert, G. Nyström, T. Zimmermann, G. Siqueira, *Carbohydr Polym* **2021**, *259*.

[6] T. Swift, L. Swanson, M. Geoghegan, S. Rimmer, *Soft Matter* **2016**, *12*, 2542.